\documentclass[12pt]{article}
\usepackage{ragged2e}
\usepackage[utf8]{inputenc}
\usepackage[colorlinks=true,citecolor=blue]{hyperref}
\usepackage[dvipsnames]{xcolor}
\usepackage{enumitem}
\usepackage{multirow}
\usepackage{pifont}
\usepackage{graphicx}
\usepackage{booktabs,arydshln}
\usepackage{algpseudocode}
\usepackage{amsmath}
\usepackage{amsfonts}
\usepackage{bbm}
\usepackage{newtxmath}
\usepackage{geometry}
\usepackage[ruled,vlined]{algorithm2e}

\usepackage{caption}
\usepackage{rotating}
\usepackage{blindtext}
\usepackage{titlesec}

\usepackage{natbib}
\usepackage{breakcites}

\newcommand{\vect}[1]{\boldsymbol{#1}}

\title{Probabilistic Crop Yields Forecasts With Spatio-Temporal Conditional Copula Using Extreme Weather Covariates}
\author{Marie Michaelides$^{*}$, Mélina Mailhot, Yongkun Li \\\\
        \small  Department of Mathematics and Statistics, Concordia University, Montréal, Canada \\
        \small $^{*}$Corresponding author \tt{marie.michaelides@concordia.ca}}
\date{2025}

\begin{document}

\maketitle

\begin{abstract}
    We introduce a novel forecasting model for crop yields that explicitly accounts for spatio-temporal dependence and the influence of extreme weather and climatic events. Our approach combines Bayesian Structural Time Series for modeling marginal crop yields, ensuring a more robust quantification of uncertainty given the typically short historical records. To capture dynamic dependencies between regions, we develop a time-varying conditional copula model, where the copula parameter evolves over time as a function of its previous lag and extreme weather covariates. Unlike traditional approaches that treat climatic factors as fixed inputs, we incorporate dynamic Generalized Extreme Value models to characterize extreme weather events, enabling a more accurate reflection of their impact on crop yields. Furthermore, to ensure scalability for large-scale applications, we build on the existing Partitioning Around Medoids clustering algorithm and introduce a novel dissimilarity measure that integrates both spatial and copula-based dependence, enabling an effective reduction of the dimensionality in the dependence structure.
\end{abstract}

\section{Introduction}

Accurate forecasting of crop yields is critical for ensuring food security, guiding agricultural management, and supporting risk assessment in the insurance industry. These forecasts are essential not only for optimizing farm-level decision-making but also for developing policies to address potential food shortages and mitigate economic risks. However, the increasing variability and intensity of climatic events, driven by climate change, have added significant complexity to this task. Unprecedented weather extremes such as droughts, floods, and heatwaves have profound impacts on agricultural systems, amplifying uncertainty in crop production. Addressing this challenge is crucial for achieving several of the United Nations' Sustainable Development Goals (SDGs), laid out in the 2030 Agenda for Sustainable Development (\cite{UnitedNations2015}), including SDG 2 (Zero Hunger), SDG 13 (Climate Action), and SDG 12 (Responsible Consumption and Production).

A key limitation of traditional yield forecasting approaches is that they model crop yields independently across regions or time periods, ignoring the spatio-temporal dependencies that are inherent in agricultural systems. The regression-based framework proposed in \cite{schlenker2009}, for example, treats locations independently and may underestimate systemic or correlated risks across regions. In practice, crop yields in different areas are often influenced by shared climatic events, such as regional droughts or heatwaves, and by temporal autocorrelation driven by persistent environmental or agronomic conditions. Failing to account for these dependencies can lead to inaccurate forecasts and underestimation of systemic risk—particularly in applications such as insurance pricing or regional risk aggregation. 

While some crop yield forecasting methods do incorporate temporal dependence—notably through ARIMA or related time series models (e.g., \cite{reddy2020})—they typically assume spatial independence and are not designed to model how yields in different regions co-evolve under shared climatic influences. Moreover, most existing techniques do not adequately capture uncertainty from extreme weather events, which are increasingly relevant under climate change. As highlighted in recent reviews (\cite{lobell2011, Ray2015, Celis2024}), more flexible and integrated approaches are needed to quantify and forecast the effects of evolving climate risks on agricultural production.

To overcome these limitations, we develop a novel forecasting framework that integrates advanced time series modeling, dependence modeling, and clustering techniques. The marginal distributions of crop yields are modeled using Bayesian Structural Time Series (BSTS), a probabilistic framework that captures trend, seasonality, and meteorological covariates while offering robust uncertainty quantification. Unlike classical ARIMA or ARIMA-GARCH models, BSTS is particularly well-suited for datasets with short historical records, a common challenge in agricultural forecasting. Its Bayesian formulation facilitates the incorporation of prior knowledge, allowing for greater interpretability and adaptability to changing climatic conditions. This approach provides a more robust alternative to traditional methods, particularly in situations where data availability is limited. The BSTS framework was introduced by \cite{scott2014} as a flexible alternative to standard time series models, making it especially valuable in scenarios where uncertainty quantification is critical.

In addition to modeling the marginal distributions, we account for the spatio-temporal dependencies in crop yields by constructing a time-varying conditional copula model. Unlike traditional copula models that assume static dependence structures, our approach allows the copula dependence parameter to evolve dynamically as a function of its past values and extreme weather covariates. This flexibility enables the model to capture shifts in dependence relationships between different regions as climatic conditions change. Copulas, first introduced by \cite{sklar1959} and extensively discussed in \cite{nelsen2006}, provide a powerful framework for modeling dependencies separately from marginal distributions, making them particularly useful for capturing complex relationships in agricultural data. The use of conditional copulas in time series settings was pioneered by \cite{patton2006}, who demonstrated how dependence structures can be made time-varying by linking copula parameters to external covariates.

A key innovation in our framework is the incorporation of dynamic Generalized Extreme Value (GEV) models to account for the role of extreme climatic events. Instead of treating climatic variables as fixed inputs, we model their temporal evolution, allowing for a more accurate representation of how extreme weather patterns influence crop yields. This approach ensures that both the magnitude and persistence of extreme events, such as prolonged droughts or heavy rainfall, are captured within the forecasting model. The GEV distribution, first introduced by \cite{Gumbel1958} in the context of extreme value theory and further formalized in \cite{Coles2001}, provides a natural framework for modeling extreme climatic risks. By integrating dynamic GEV outputs as exogenous variables in both the BSTS model and the copula dependence structure, we not only quantify the uncertainty surrounding extreme weather effects but also enhance the ability to assess climate-related risks.

Another major challenge in modeling crop yields across multiple regions is the high dimensionality of the dependence structure. To address this, we use a clustering approach based on Partitioning Around Medoids (PAM), which allows us to reduce the dimensionality while preserving key regional patterns. Clustering regions when working with extreme weather events is crucial, as demonstrated in the recent work of \cite{boulin2025}, where the authors apply a clustering framework to identify subregions with distinct extremal dependence structures in compound precipitation and wind speed events. Our clustering framework is based on a novel dissimilarity measure that combines both spatial and copula-based dependence metrics. Unlike previous approaches that rely on empirical copulas, such as the one presented in \cite{palacios2023}, we derive our dissimilarity measure from Kendall’s tau, ensuring greater interpretability and statistical rigor. By grouping regions based on this combined dissimilarity, we reduce computational complexity while maintaining a meaningful representation of dependence structures, making our model scalable for large datasets.

In addition to introducing this novel forecasting methodology, we present the full model fitting and forecasting algorithms required for practical implementation. Beyond theoretical developments, we provide a detailed algorithmic framework for both estimation and prediction, ensuring that our methodology can be readily applied to real-world data. To facilitate reproducibility and practical adoption, we have also developed an R package implementing our methods. This package enables users to easily replicate our results, apply the model to their own datasets, and extend the methodology as needed. 

By combining BSTS for marginal modeling, a time-varying copula for spatio-temporal dependence, dynamic GEV for extreme weather uncertainty, and a PAM-based clustering strategy for scalability, this paper presents a comprehensive, interpretable, and scalable solution for crop yield forecasting under evolving climatic risks. The remainder of this paper is structured as follows. We begin in Section \ref{sec:literature} with a review of the relevant literature, focusing on existing crop yield forecasting methods and dependence modeling techniques. Next, in Section \ref{sec:data}, we describe the dataset used for our analysis, which includes winter wheat yield data from Ontario, Canada, and the 28 Core Climate Extreme Indices developed by the Expert Team on Climate Change Detection And Indices. We then introduce our forecasting framework in Sections \ref{sec:model} and \ref{sec:clustering}, detailing the modeling components and the clustering methodology. This is followed by an empirical application of our model to the Canadian dataset in Section \ref{sec:application}, where we demonstrate its predictive performance and scalability. Finally, Section \ref{sec:conclusion} concludes with a discussion of our findings and potential avenues for future research.

\section{Literature Review}
\label{sec:literature}

Traditional crop yield prediction models often rely on simplistic statistical methods that fail to account for the complex interplay between climatic factors, spatial variability, and crop performance. These models typically assume stationarity and overlook the dynamic nature of climatic influences, which are increasingly volatile in the context of global climate change. Recent works, such as that of \cite{feng2023method}, highlight the importance of incorporating more complex climatic factors and spatial variability in crop yield predictive models by providing a comprehensive review of methodologies assessing the impact of climate change on agricultural outcomes. Advancements in statistical modeling have sought to address these limitations by integrating spatio-temporal dependencies and external climate risk factors, providing a more comprehensive framework for understanding and predicting agricultural productivity. For instance, \cite{boyer2014} explore the skewness in crop yield distributions and extend existing models to incorporate skew-normal distributions, emphasizing the need for methods that reflect the realities of agricultural production. Similarly, \cite{shen2017} discuss adaptive local parametric estimation methods that address non-stationarity in crop yield data due to technological and climatic changes.

Efforts to model spatio-temporal dependence in crop yields have increasingly turned to advanced statistical frameworks, such as spatial regression models and hierarchical Bayesian approaches. For example, \cite{veron2015} demonstrate the importance of spatially explicit models in analyzing crop yields in the Pampas region of Argentina, highlighting the significant impacts of precipitation and temperature variability. Similarly, \cite{rosenzweig2014} assess the impacts of climate variability and change on global crop productivity using simulation models that incorporate spatially explicit climate scenarios and crop growth processes. \cite{kim2019} examine the effects of heavy rainfall events on maize yield trends, indicating that climate variability can lead to ambiguous yield responses. These findings align with those of \cite{chemere2018}, who employ ARIMAX models to evaluate the influence of climatic factors on maize yields in Korea. The actuarial literature also underscores the importance of accurate crop yield modeling, particularly for agricultural insurance. Yield forecasts are essential for setting fair premium rates and managing risks associated with climate variability. Studies such as \cite{zhu2018} and \cite{cheung2024} incorporate dynamic factors and extreme weather indices into insurance frameworks, emphasizing the need for statistical models that capture the joint impacts of climate variability and yield dynamics. However, while these approaches capture important aspects of crop yield variability, they often fall short in modeling the joint dependence structure of crop yields across regions or time periods, particularly when extreme events or tail dependencies are of interest. Addressing this gap requires a more flexible framework capable of capturing both marginal dynamics and joint dependencies.

One promising solution lies in copula-based methods, which are well-suited to capturing joint dependence structures. For instance, \cite{salvadori2014} use copulas to model the joint distribution of extreme rainfall and river flow, while \cite{alidoost2019} employ copulas to investigate the effects of climate extremes on crop yield and price dynamics. Similarly, \cite{wang2019} develop a copula-based drought index to assess agricultural drought impacts, emphasizing the importance of modeling dependencies between climatic factors and crop yields. While these applications underscore the versatility of copulas, most of these approaches assume static dependencies, which may not fully reflect the dynamic nature of agricultural systems under changing climatic and economic conditions.

To address this limitation, conditional time-varying copula models explicitly incorporate the temporal evolution of dependencies while allowing for external covariates. First introduced by \cite{patton2006} in the context of exchange rate modeling, these models have since gained traction in fields such as environmental sciences and agriculture. For instance, \cite{nasri2019copula} propose a theoretical framework for dynamic copula-based models, focusing on multivariate financial time series with regime-switching copulas. Despite their potential, conditional time-varying copulas remain underutilized in agricultural applications. Integrating these copulas with advanced time series models for the marginals could provide a more holistic approach to model spatio-temporal dependencies in crop yields.

Incorporating time-series methods into marginal modeling is crucial for capturing the temporal evolution of crop yields. While traditional approaches like ARIMA or ARIMAX have been widely used, they often fail to adequately quantify uncertainty, particularly when data are limited. In contrast, Bayesian Structural Time Series models provide a robust alternative. Since their introduction by \cite{scott2014}, they have been applied in a wide variety of fields such as in finance to forecast stock prices (\cite{Katarina2023}) or in biomedical studies for sensor data analysis (\cite{Liu2021}). These models integrate state-space structures with Bayesian inference, enabling the decomposition of time series into components such as trend, seasonality, and covariate effects. Moreover, BSTS models dynamically include covariates and quantify the uncertainty inherent in agricultural data, making them particularly well suited for estimating the pseudo-residuals used in conditional copula models. 

Building on this foundation, our model focuses specifically on extreme climate events—such as heavy rainfall, heatwaves, and droughts—as key risk factors influencing crop yields. Rather than including a wide array of meteorological variables, we show that modeling extremes alone can sufficiently capture the relevant climatic risks affecting yield variability. To achieve this, we use Generalized Extreme Value (GEV) distributions, which are well-suited for characterizing rare but impactful events. Traditional GEV models assume stationarity, which may not hold in a changing climate. We instead use dynamic GEV models, which allow parameters such as location, scale, and shape to evolve over time or with covariates (\cite{gilleland2016}). For instance, \cite{fischer2015} apply dynamic GEV models to investigate the rising frequency of heatwaves, while \cite{lobell2011} explore how extreme temperature and precipitation events affect global crop yields.

By integrating conditional time-varying copulas, BSTS models, and dynamic GEV distributions, we provide in this paper a unified framework for modeling crop yields under climate risk. Capturing the temporal evolution of marginal distributions, dependence structures, and external risk factors, our approach addresses key limitations of existing models and offers robust tools for forecasting agricultural outcomes in the context of climate change. 


\section{Data}
\label{sec:data}
In this section, we present and explore the data that we will use in the remainder of the paper to build our model and analyze the impacts of extreme climate events on crop yields.

\subsection{Crops data}
\label{sec:cropdata}
Although our method can be extended to a sparse domain, we focus our analysis on public data related to crop yields in the Canadian province of Ontario. The province is divided into five Census Agricultural Regions (CARs), which are defined by Statistics Canada to group census divisions with similar agricultural characteristics. These regions are commonly used for statistical reporting and analysis of agricultural trends, as they provide a coherent framework for comparing agricultural practices and results in different areas within the province.

Each CAR is further subdivided into smaller census divisions that serve as the primary administrative units for collecting and reporting agricultural data. Census divisions typically align with county or regional boundaries and provide granular information on agricultural production, demographics, and land use. The crop yield data used in this study are available at the level of these census divisions, offering a detailed view of agricultural productivity across Ontario.

As illustrated in Figure \ref{fig:map}, we start by considering Census Agricultural Regions 1 to 4, which include Southern Ontario, Western Ontario, Central Ontario, and Eastern Ontario. These regions are located in the southern and southeastern parts of the province, where agriculture is more prevalent due to favorable climate and soil conditions. 
This regional focus ensures that the study remains relevant in areas where agriculture plays a significant role in the local economy and land use, while also allowing us to take advantage of the availability of detailed public data on crop yields from these divisions.

\begin{figure}[h!]
\centering
\includegraphics[width=.8\textwidth]{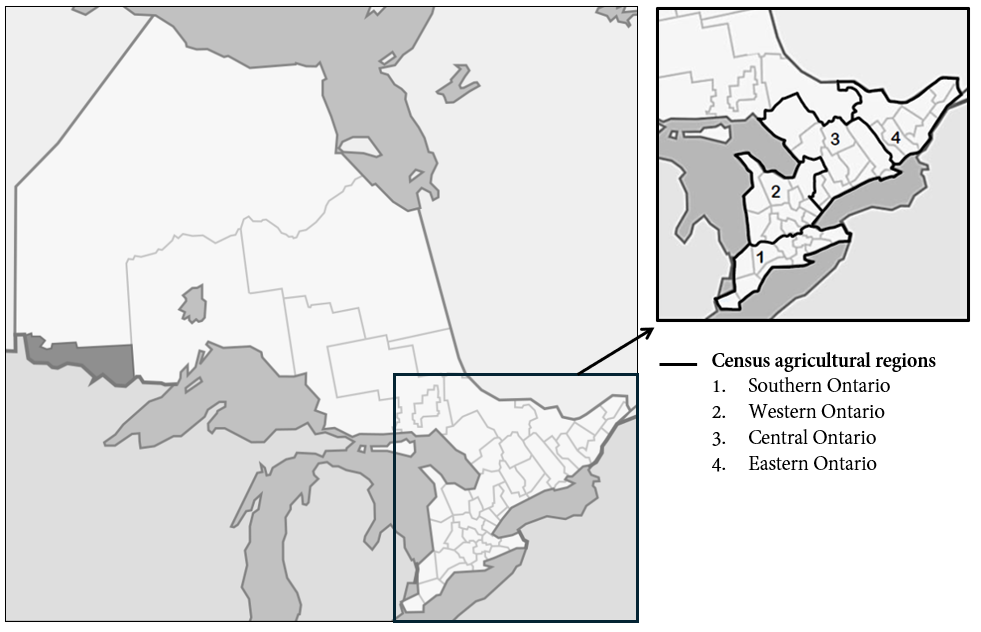}
\caption{Census agricultural regions of Ontario.}
\label{fig:map}
\end{figure}

More specifically, we collect winter wheat crop yield data, spanning the years 1950 to 2022. These data were sourced from the official Ontario data portal\footnote{https://data.ontario.ca/}, which provides a reliable and detailed historical record of agricultural production across the province. The dataset includes annual yield measurements of a large variety of crops, including winter wheat. 

Although public data are available for most census divisions in Ontario, the completeness and quality of the data vary across divisions. In some cases, data for certain years are missing, limiting their usability for longitudinal analysis. To ensure robustness and consistency in our study, we focus on twenty-four specific census divisions with complete data records. 



\subsection{Core Climate Extreme Indices}
In addition to crop yields data, we collect climatic and weather-related data for each census division presented in Section \ref{sec:cropdata}. More specifically, we use the Core Climate Extreme Indices developed by the Expert Team on Climate Change Detection and Indices (ETCCDI). These indices are standardized methods to objectively measure and characterize climate change, offering measures that are pivotal for climate change detection. We use them as a framework to assess the impact of extreme climate conditions on wheat crop yields. Given that our crop yields data are at the more granular level of the census divisions, the selected extreme climate indices must be compatible with this geographical resolution. Accordingly, we use data from 28 extreme climate indices provided by ClimateData.ca\footnote{https://climatedata.ca/}, chosen for their accessibility, objectivity, and applicability. This data portal was launched in June 2019 by the Government of Canada in an effort to provide projected and historical climate data openly and freely to researchers, public health professionals, engineers, planners and anyone who require more detailed information on climate change. The 28 indices are presented in Table \ref{tab:etccdi_indices} of Appendix \ref{app:ETCCDI}.

Given the limited length of the crop yield time series (1950–2022), including all 28 indices would result in over-parameterization. To reduce dimensionality, we first assess linear correlation among the indices, which reveals strong collinearity within temperature- and precipitation-based subgroups (e.g., hot days and tropical nights, or various precipitation duration metrics). This analysis is visualized in the correlation heatmap in Appendix \ref{app:1} (Figure \ref{fig:map3}). We also explore tail dependence (Figures \ref{fig:map§} and \ref{fig:map4}) to ensure that selected indices capture relationships in the extremes—a key aspect for understanding climate risk. Notably, strong upper tail dependence is observed among temperature-based indices, particularly between hot days and mean temperatures.

Based on both collinearity and tail dependence, we retain five indices that best represent the main climatic drivers without redundancy:
\begin{enumerate} 
    \item \texttt{cdd}: Maximum number of consecutive dry days, capturing drought risk. 
    \item \texttt{rx1day}: Annual maximum 1-day precipitation, relevant for assessing the impact of extreme rainfall. 
    \item \texttt{frost\_days}: Number of frost days, reflecting potential cold stress during the growing season. 
    \item \texttt{tg\_mean}: Mean daily temperature, chosen as a representative of overall thermal conditions. 
    \item \texttt{txgt\_25}: Number of days exceeding 25$^{\circ}$C, reflecting heat stress and linked to other hot day metrics. 
\end{enumerate}

\begin{figure}[h!]
\centering
\includegraphics[width=1\textwidth]{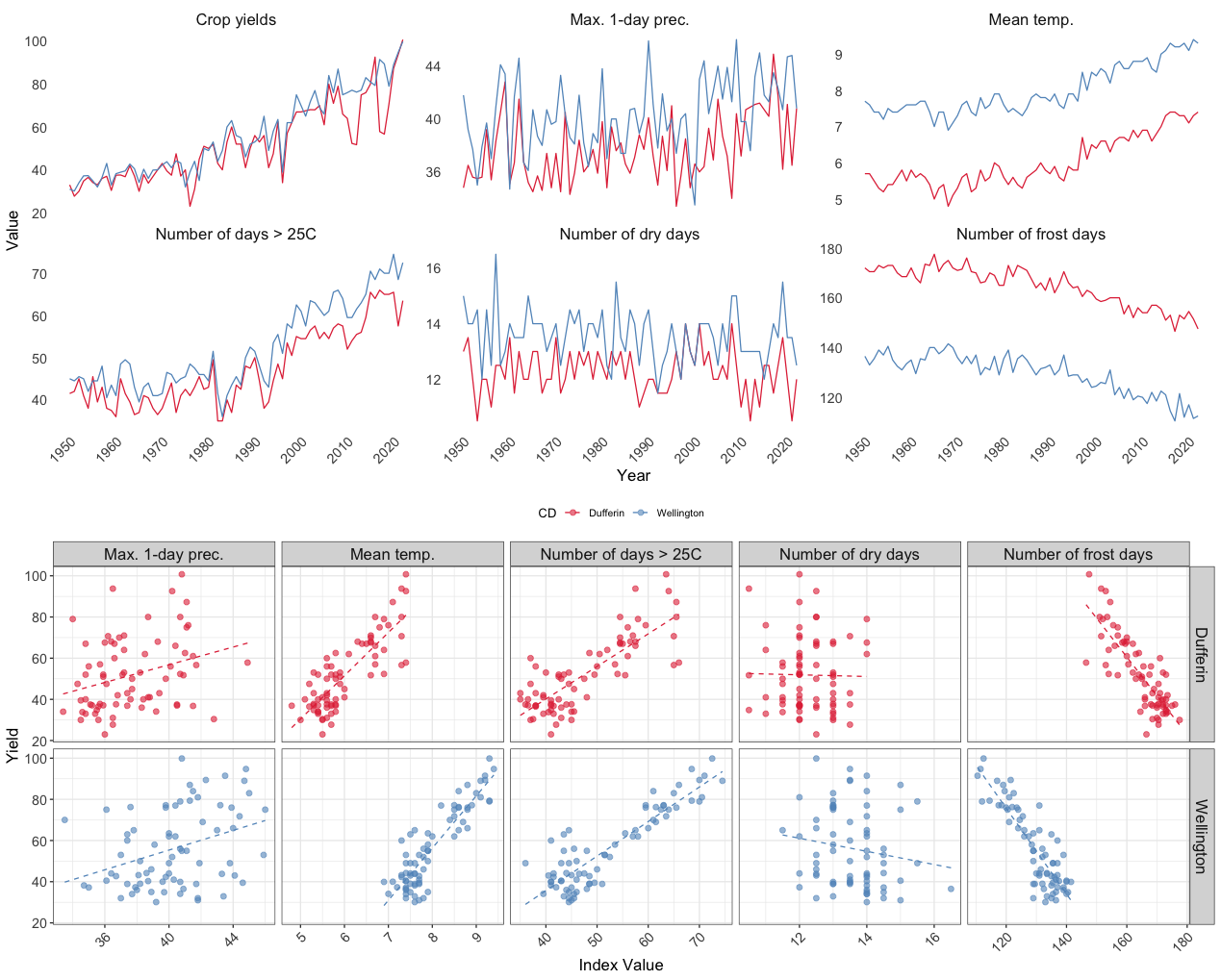}
\caption{Time series plots and scatter plots of indices.}
\label{fig:map5}
\end{figure}

Figure \ref{fig:map5} presents time series and scatter plots of these selected indices for the census divisions of Dufferin (in red) and Wellington (in blue). The plots show upward trends in temperature-related indices and yield, consistent with warming climate patterns and improved agricultural productivity, while precipitation indices show no clear trend. Linear relationships are more pronounced between yield and temperature-related indices than with precipitation-related ones, supporting the choice of variables retained in the model.

\section{Statistical Model}
This section introduces the forecasting framework for crop yields. We begin by presenting the notation that will be used throughout the section. We then introduce the marginal time series models for crop yields data, namely the Bayesian Structural Time Series Models. We then show how we build the time-varying conditional copula models on the fitted residuals of the BSTS, and how we incorporate extreme weather effects in the model.
\label{sec:model}
\subsection{Notation}
We first describe the notation that we will use throughout the description of our model. In particular, we introduce the following components:
\begin{itemize}
    \item $Y_{d,t} \in \mathbb{R}$ is a continuous random variable that denotes the crop yield in region $d$ at time $t$, for $d=1,\ldots,D$ and $t=1,\ldots,T$.
    \item $\vect{X}_{t}$ is a $1\times M$ vector of climatic or meteorological random variables measured at time $t$ across all $D$ regions. 
    \item $\vect{Z}_{d,t}$ is a $1\times M$ vector of climatic or meteorological random variables measured at time $t$ in region $d$.
\end{itemize}
As described in Section \ref{sec:data}, we use the 28 core climate extreme indices defined by the Expert Team on Climate Change Detection and Indices (ETCCDI) as variables $\vect{X}_t$ and $\vect{Z}_{d,t}$. The risk factors $\vect{Z}_{d,t}$ thereby describe these indices in each region $d$, while $\vect{X}_{t}$ measures them across all $D$ regions under consideration in our analysis.



\subsection{Marginal Time Series Models for Crop Yields}
\label{sec:TS}
We present here the marginal models employed to fit crop yields in each region, using a time series framework as outlined in Section \ref{sec:data}. While traditional time series models like ARIMAX-GARCH may seem appealing, we discuss in this section the reasons why they may not be suitable for crop yield data due to the limited number of observations, a common challenge compared to typical financial time series. Instead, we propose Bayesian Structural Time Series models as a more appropriate alternative.

A Bayesian Structural Time Series assumes that the observations of a given time series are determined by a hidden state that evolves over time according to a Markov process. Bayesian Structural Time Series models offer a highly flexible framework capable of accommodating a wide variety of components to suit different data structures and temporal patterns. In this case, we propose a BSTS model that incorporates a local linear trend, a seasonal component, an autoregressive (AR) component, and a dynamic regression component, seeking this level of flexibility to properly account for the specific nature and dependence patterns of crop yield data. The observation equation is given by
\begin{align}
\label{eq:BSTS}
    y_{d,t} = l_{d,t} + s_{d,t} + \sum_{l=1}^{p} \psi_{d,l} e_{d,t-l} + \sum_{m=1}^{M}\beta_{d,m,t} Z_{d,m,t} + \epsilon_{d,t},
\end{align}
where $\epsilon_{d,t}\sim \mathcal{N}(0,\sigma_{\epsilon_d}^2)$. We describe the different components of this equation below.
\begin{itemize}
    \item The local linear trend is represented by $l_{d,t}$ and modeled by a random walk such that
    \begin{align*}
        l_{d,t+1} &= l_{d,t} + \tau_{d,t} + \nu_{d,t}, \hspace{0.3cm} \nu_{d,t} \sim \mathcal{N}(0,\sigma_{\nu_d}^2),
    \end{align*}
    where $l_{d,t}$ stands for the local level of the time series, while $\tau_{d,t}$ represents its slope (or trend). This trend typically evolves as
    \begin{align*}
        \tau_{d,t+1} = \tau_{d,t} + \zeta_{d,t}, \hspace{0.3cm} \zeta_{d,t} \sim \mathcal{N}(0,\sigma_{\zeta_d}^2).
    \end{align*}
    These combined terms capture the linear trend in crop yield data, allowing for gradual change over time.
    \item When applicable, $s_{d,t}$ is the seasonal component, accounting for recurring patterns or seasonal effects. It is computed at each time step as
    \begin{align*}
        s_{d,t} &= - \sum_{i=1}^{S-1}s_{d,t-i} + \omega_{d,t}, \hspace{0.3cm} \omega_{d,t} \sim \mathcal{N}(0,\sigma_{\omega_d}^2),
    \end{align*}
    where $S$ denotes the periodicity observed in the data. Note that if we work with yearly crop yields, $S$ will be equal to 1, thereby removing the seasonal component from our observation equation.
    \item The coefficients $\psi_{d,l}$, for $l=1,\ldots,p$ capture the effects of previous lags of the residuals. This auto-regressive component models the serial correlation in the crop yields, once the trends, seasonality and other components have already been accounted for. This AR(p) process is defined as
    \begin{align*}
        e_{d,t} = \sum_{l=1}^{p}\psi_{d,l} e_{d,t-l} + \eta_{d,t}, \hspace{0.3cm} \eta_{d,t} \sim \mathcal{N}(0,\sigma_{\eta_d}^2)
    \end{align*}
    By explicitly modeling the dependence structure in the residuals, the AR process ensures that any temporal correlation remaining after the contributions of trends, seasonality, and risk factors is effectively removed, leaving the residuals as uncorrelated white noise.
    \item We can incorporate exogeneous risk factors in the BSTS model by capturing their impact through the dynamic coefficients $\vect{\beta}_{d,t}$. This dynamic regression approach differs from static regression by allowing the coefficients to evolve over time, thereby adapting to changing relationships between covariates and the response variable, which is particularly important in dynamic systems like crop yields influenced by environmental factors. For $d=1,\ldots,D$, $m=1,\ldots,M$ and $t=1,\ldots,T$, these dynamic regression coefficients are given by
    \begin{align*}
       \beta_{d,m,t+1} = \beta_{d,m,t} + \lambda_{d,t},\hspace{0.3cm} \lambda_{d,t} \sim \mathcal{N}(0,\sigma_{\lambda_d}^2).
    \end{align*}
    \item The remaining random fluctuations or unexplained variability in the crop yield data, not accounted for by the trend, seasonal, auto-regressive or exogenous components, are captured by the observation noise $\epsilon_{d,t}$.
\end{itemize}

BSTS models offer several advantages over classic time series approaches such as the commonly used ARIMAX-GARCH framework. In particular, they are a better choice when dealing with time series with few data points like the crop yields data described in Section \ref{sec:data}. BSTS models are specifically designed to handle small datasets by incorporating prior information, which helps stabilize parameter estimation and reduces overfitting. Unlike ARIMAX-GARCH models that rely heavily on a large number of observations for accurate parameter estimation, BSTS models leverage Bayesian inference to combine prior knowledge with observed data, resulting in more robust estimates even when the data is sparse.

We write the density function for the BSTS described above as
\begin{align*}
    f(y_{d,t}\mid \vect{\psi}_d, \sigma_{\epsilon_d}, \sigma_{\nu_d},\sigma_{\zeta_d},\sigma_{\omega_d}, \sigma_{\eta_d}, \sigma_{\lambda_d}) = \frac{1}{\sqrt{2\pi\sigma^2_{\epsilon_d}}}\exp \Bigg(-\frac{\epsilon_{d,t}^2}{2\sigma^2_{\epsilon_d}}\Bigg),
\end{align*}
where
\begin{align*}
    \epsilon_{d,t} = y_{d,t} - l_{d,t} - s_{d,t} - \sum_{l=1}^{p}\psi_{d,l} e_{d,t-l} - \sum_{m=1}^{M}\beta_{d,m,t} Z_{d,m,t}.
\end{align*}
We then assume prior distributions for the model parameters $\vect{\psi}_d$ and $\vect{\sigma}^2_d = \{\sigma^2_{\epsilon_d}, \sigma^2_{\nu_d}, \sigma^2_{\zeta_d}, \sigma^2_{\omega_d}, \sigma^2_{\eta_d},\sigma^2_{\lambda_d}\}$. We retrieve the posterior distribution by means, for example, of Markov Chain Monte Carlo simulations:
\begin{align*}
    f_{\text{BSTS}}(\vect{\psi}_d, \vect{\sigma}^2_d ) \propto \prod_{t=1}^{T}& f(\vect{y}_{d}\mid \vect{\psi}_d, \vect{\sigma}^2_d) \pi(\vect{\psi}_d)\pi(\vect{\sigma}^2_d),
\end{align*}
where $\pi(\vect{\sigma}^2_d)=\pi(\sigma_{\epsilon_d})\pi(\sigma_{\nu_d})\pi(\sigma_{\zeta_d})\pi(\sigma_{\omega_d})\pi(\sigma_{\eta_d})\pi(\sigma_{\lambda_d})$. The distributions $\pi(.)$ are the priors selected for the different model parameters. The likelihood function of the BSTS models for the crop fields of each region $d$ is defined as
\begin{equation}
\label{eq:LBSTS}
\begin{aligned}
    \mathcal{L}_{\text{BSTS}}(\vect{\psi},\vect{\sigma}^2) &= \prod_{d=1}^{D} f_{\text{BSTS}}(\vect{\psi}_d, \vect{\sigma}^2_d ) \\
    &\propto f(\vect{y}_{d}\mid \vect{\psi}_d, \vect{\sigma}^2_d) \pi(\vect{\psi}_d)\pi(\vect{\sigma}^2_d).
\end{aligned}
\end{equation}


\subsection{Time-varying conditional copula}
\label{sec:copula}
We now describe how we capture the spatio-temporal dependence in crop yield data using a time-varying conditional copula. We write the conditional joint cumulative distribution function of the crop yields across all $D$ regions as
\begin{align*}
    F(y_{1,t},...,y_{D,t} \mid \vect{X}_t)  = C(u_{1,t},\ldots,u_{D,t}\mid\theta_t(\vect{X}_t)).
\end{align*}
In this expression, $C(.)$ is the copula used to jointly model the crop yields across all regions. The terms $u_{d,t}$ for $d=1,\ldots,D$ are pseudo-observations retrieved from the BSTS models fitted for the crop yields in each region $d$. If these marginal models have been correctly fitted, then the residuals $\epsilon_{d,t}$ are independent for all $t=1,\ldots,T$, implying the independence of the pseudo-observations as well. The copula parameter, $\theta_t(\vect{X}_t)$, is the copula parameter varying with time and with the vector of risk factors $\vect{X}_t$. The copula parameter varies over time to account for temporal changes in the dependence structure. To fully capture any temporal dependence influencing the crop yields, we model $\theta_t(\vect{X}_t)$ using the following expression:
\begin{align}
\label{eq:theta}
    \theta_t(\vect{X}_t) = g\Big(\omega + \alpha \theta_{t-1} + \sum_{m=1}^{M}\gamma_{k,t} X_{k,t}\Big).
\end{align}
This represents the time-varying copula parameter as a function of both its past values and external covariates. The term $\omega$ denotes the intercept, capturing the baseline level of $\theta_t$. The coefficient $\alpha$ quantifies the influence of the previous lag, $\theta_{t-1}$, reflecting how the dependence structure evolves over time based on past values. The coefficients $\vect{\gamma}_t$ capture the effects of climatic or meteorological variables on the copula parameter. The function $g(.)$ ensures that $\theta_t$ remains within an appropriate range for the chosen copula family.

The pseudo-likelihood function for the conditional copula model is then given by
\begin{align}
\label{eq:Lcop}
    \mathcal{L}_{\text{copula}}(\vect{\omega},\vect{\alpha},\vect{\gamma}) = \prod_{t=1}^{T}c(u_{1,t},\ldots,u_{D,t}\mid \theta_t(\vect{X}_t)) ,
\end{align}
where $c(.)$ is the copula density function.

\subsection{Incorporating extreme weather effects}
\label{sec:GEV}
The weather and meteorological variables introduced in Section \ref{sec:data} are included in both Equation (\ref{eq:BSTS}) for the marginal models of crop yields and Equation (\ref{eq:theta}) for the time-varying copula parameter. To incorporate the effects of extreme climate events into our model, a compelling approach is to analyze these variables using extreme value theory. Specifically, we focus on modeling the annual maximum values of each variable $t=1,\ldots,T$ using Generalized Extreme Value (GEV) distributions. This allows us to effectively capture the impact of extreme events on both the marginal distributions and the dependence structure of crop yields.

In addition, to account for the time-variation in the covariates and to align with the dynamic structure used in the BSTS model, we model the location parameter $\mu_t$ of the GEV distribution as an autoregressive process of order 1 (AR(1)). This allows us to capture both short-term fluctuations and long-term dependencies in the covariates. We thereby work with a dynamic GEV distribution whose cumulative distribution function for $Z_{d,m}$ is given by

\begin{align*}
    F_{Z_{d,m}}(z_{d,m,t} \mid \mu_{d,m,t}, \sigma_{d,m},\xi_{d,m}) = \exp \Bigg\{ -\Big[1 + \xi_{d,m} \frac{z_{d,m,t}-\mu_{d,m,t}}{\sigma_{d,m}}\Big]^{-1/\xi_{d,m}}\Bigg\}, 
\end{align*}
where $\sigma_{d,m}$ and $\xi_{d,m}$ are, respectively, the scale and shape parameters of risk factor $k$ in region $d$. We let the location parameter $\mu_{d,m,t}$ be time-varying and model it as
\begin{align*}
    \mu_{d,m,t} = \phi_{d,m} \mu_{d,m,t-1} + \epsilon_{d,m,t}, 
\end{align*}
with $\epsilon_{d,m,t} \sim \mathcal{N}(0,\sigma^2_{\mu_{d,m}})$. In this expression, $\phi_{d,m}$ is the autoregrssive coefficient, capturing the persistence $\mu_{d,m,t}$ over time, and $\sigma_{\mu_{d,m}}^2$ is the variance of the innovations $\epsilon_{d,m,t}$, controlling the magnitude of stochastic variations in $\mu_{d,m,t}$. The density function associated to this AR(1) model can be written as
\begin{align*}
    f_{\mu_{d,m}}(\mu_{d,m,t}) = \frac{1}{\sqrt{2\pi}\sigma_{\mu_{d,m}}}\exp\Big(-\frac{(\mu_{d,m,t} - \phi_{d,m}\mu_{d,m,t-1})^2}{2\sigma^2_{\mu_{d,m}}}\Big).
\end{align*}

This dynamic specification for $\mu_{d,m,t}$ aligns with the time-series framework of the BSTS model, which similarly captures temporal evolution in the response variable and its covariates. By modeling $\mu_{d,m,t}$ as an AR(1) process, we allow the location parameter of the GEV distribution to evolve over time, reflecting the inherent temporal variability in extreme covariate behavior. This structure introduces persistence in $\mu_{d,m,t}$ while enabling it to adapt to new observations, making the model more flexible and realistic for time-dependent data.

Note that in this paper, we choose to keep the other GEV parameters, namely $\sigma_{d,m}$ and $\xi_{d,m}$, fixed. This choice is motivated by both practical and interpretability considerations. First, the location parameter governs the central tendency of extremes and is directly influenced by systematic changes in climatic conditions. Allowing it to vary captures shifts in the level of extreme events (e.g., increasing maximum temperatures or precipitation). While allowing the other parameters to vary as well could capture changing variability of extremes, doing so introduces additional complexity and increases the number of parameters to estimate, which may not be feasible given our relatively short time series. Such arguments have similarly been discussed, among others, in the works of \cite{huerta2007} and \cite{depaola2018}.

The probability density function of the GEV is then written as
\begin{equation}
\begin{aligned}
\label{eq:GEV}
    f_{Z_{d,m}}(z_{d,m,t}\mid \mu_{d,m,t}, \sigma_{d,m},\xi_{d,m}) &= \frac{1}{\sigma_{d,m}}\Bigg[1 + \xi_{d,m} \frac{z_{d,m,t}-\mu_{d,m,t}}{\sigma_{d,m}}\Bigg]^{-1-1/\xi_{d,m}} \\
    &\hspace{2cm} \times \exp\Bigg\{-\Big[1 + \xi_{d,m} \frac{z_{d,m,t}-\mu_{d,m,t}}{\sigma_{d,m}}\Big]^{-1/\xi_{d,m}}\Bigg\}.
\end{aligned}
\end{equation}

The likelihood contribution of the dynamic GEV models in the full model likelihood is given by
\begin{equation}
\begin{aligned}
\label{eq:LGEV}
    \mathcal{L}_{GEV}(\vect{\phi},\vect{\sigma_{\mu}},\vect{\sigma},\vect{\xi}) &= \prod_{t=1}^{T} \prod_{d=1}^{D} \prod_{k=1}^{M} \Bigg( \frac{1}{\sigma_{d,m}}\Bigg[1 + \xi_{d,m} \frac{z_{d,m,t}-\mu_{d,m,t}}{\sigma_{d,m}}\Bigg]^{-1-1/\xi_{d,m}}\\
    &\hspace{3cm} \times \exp\Bigg\{-\Big[1 + \xi_{d,m} \frac{z_{d,m,t}-\mu_{d,m,t}}{\sigma_{d,m}}\Big]^{-1/\xi_{d,m}}\Bigg\}  \\
    & \hspace{4cm} \times \frac{1}{\sqrt{2\pi}\sigma_{\mu_{d,m}}}\exp\Big(-\frac{(\mu_{d,m,t} - \phi_{d,m}\mu_{d,m,t-1})^2}{2\sigma^2_{\mu_{d,m}}}\Big)\Bigg).
\end{aligned}
\end{equation}

\subsection{Full model pseudo-likelihood}
We now combine the results from Sections \ref{sec:TS}, \ref{sec:copula} and \ref{sec:GEV} to write the likelihood function of our full model. The full set of parameters is given by $\vect{\Theta} = \{\vect{\alpha}, \vect{\gamma}, \vect{\sigma}_{\theta}, \vect{\phi}, \vect{\sigma_{\mu}}, \vect{\sigma}, \vect{\xi}\}$. We write the full model pseudo-likelihood by combining the results from Equations (\ref{eq:Lcop}) and (\ref{eq:LGEV}) as
\begin{equation}
\begin{aligned}
\label{eq:fulllik}    
    \mathcal{L}_{\text{full}}(\vect{\Theta}) &= \mathcal{L}_{\text{copula}}(\vect{\omega}, \vect{\alpha}, \vect{\gamma}) \times \mathcal{L}_{\text{GEV}}(\vect{\phi}, \vect{\sigma_{\mu}}, \vect{\sigma}, \vect{\xi}) \\
    & = \prod_{t=1}^{T} \Big\{ c(u_{1,t},\ldots,u_{D,t}\mid \theta_t(\vect{X}_t)) \prod_{d=1}^{D} \prod_{k=1}^{M} \big(f_{Z_{d,m}}(z_{d,m,t}\mid \mu_{d,m,t}, \sigma_{d,m},\xi_{d,m}) f_{\mu_{d,m}}(\mu_{d,m,t})  \big) \Big\}.
\end{aligned}
\end{equation}By modeling the time-varying location parameter $\mu_{d,m,t}$ of the GEV distribution as an AR(1) process, we introduce a dynamic structure that reflects the temporal evolution of extreme covariates, consistent with the time-varying dependence captured in the copula model through $\theta_t(\vect{X}_t)$. This alignment ensures coherence in the framework, as both components incorporate temporal dependencies and allow the model to jointly account for dynamic changes in marginal covariates and their influence on the dependence structure over time.

Using GEV also allows to simplify the model and only include the likelihood contribution of each variable once since we now have for each variable $k$, with $m=1,\ldots,M$ that
\begin{align*}
        X_{m,t} = \max_{d=1,\ldots,D} Z_{d,m,t}.
\end{align*}
The maximum value of a variable across all $D$ regions will always be equal to the maximum value in one of the regions. 

An illustration of our model, using a simple example, is provided in the supplementary material.

\section{Clustering algorithm}
\label{sec:clustering}

In our proposed time-varying conditional copula framework for crop yield forecasting, we model the spatio-temporal dependence between crop yields across different census divisions while incorporating the effects of extreme climatic events. However, a significant challenge arises due to the high-dimensional nature of the copula model when the number of census divisions is large. A direct implementation of the full copula model across all census divisions would be computationally expensive and infeasible due to the curse of dimensionality.

To address this scalability issue, we propose the use of a clustering algorithm to group census divisions with similar dependence structures. Instead of estimating a single high-dimensional copula, we estimate multiple lower-dimensional copulas within each cluster. This reduces computational complexity while preserving essential dependence structures. Specifically, we adopt a Partitioning Around Medoids (PAM) algorithm similar to that proposed by \cite{palacios2023}. The PAM algorithm is a robust clustering technique that partitions census division pairs into homogeneous groups based on a dissimilarity measure. It is particularly well-suited for our application because it provides robust clustering even when using a non-Euclidean dissimilarity measure. Unlike K-means, which relies on Euclidean distances and is sensitive to outliers, PAM minimizes the total dissimilarity within each cluster by selecting representative observations called medoids. In this paper, we quantify the dissimilarity between two pairs of census divisions using a combined dissimilarity measure, incorporating both spatial distance and copula-based dependence similarity.

For each pair of census divisions $(i,j)$, with $i,j=1,\ldots,D$ and $i\neq j$, we estimate the time-varying copula parameter $\theta_t^{(i,j)}$ for each time $t=1,\ldots,T$. We then convert these parameters into Kendall’s tau values, using the known relationship for a given copula family. For example, for a Clayton copula:
\begin{equation}
\tau_t^{(i,j)} = \frac{\theta_t^{(i,j)}}{\theta_t^{(i,j)} + 2}.
\end{equation}

The copula-based dependence similarity between two pairs $(i,j)$ and $(k,l)$ is computed as the average absolute difference in their Kendall’s tau values over time:
\begin{equation}
d^{\text{copula}}_{(i,j),(k,l)} = \frac{1}{T} \sum_{t=1}^{T} \left| \tau_t^{(i,j)} - \tau_t^{(k,l)} \right|.
\end{equation}

Since the copula dissimilarity is computed for pairs of census divisions, we aggregate it into a census-level dissimilarity matrix by taking the maximum dissimilarity of all pairs containing a given census division. This means that for each census division $i$, its aggregated dissimilarity with another division $j$ is given by
\begin{align*}
    d_{i,j}^{\text{copula}} = \max_{(i,k)\in \mathcal{P}_i, (j,l)\in \mathcal{P}_j} d^{\text{copula}}_{(i,j),(k,l)},
\end{align*}
where $\mathcal{P}_i$ represents the set of all pairs involving census division $i$, and $\mathcal{P}_j$ represents the set of all pairs involving $j$. Using the maximum ensures that census divisions are assigned a dissimilarity value based on their most distinct relationships, highlighting the strongest dependencies in the data.

The spatial distance between census divisions $i$ and $j$ is denoted by $d_{(i,j)}^{\text{spatial}}$, where spatial proximity is computed using the Haversine formula, which measures the great-circle distance between two census divisions based on their latitude and longitude:

\begin{align*}
    d_{i,j}^{\text{spatial}} = R \arccos\big(\sin \psi_i \sin \psi_i + \cos \psi_i \cos \psi_j \cos(\lambda_i-\lambda_j)  \big),
\end{align*}
where $R$ denotes the Earth's radius in km, while $\psi$ and $\lambda$ represent, respectively, the latitude and longitude of census divisions $i$ and $j$.

Finally, we define the combined dissimilarity measure as:
\begin{equation}
d_{(i,j)} = \beta d_{(i,j)}^{\text{spatial}} + (1-\beta) d^{\text{copula}}_{(i,j)},
\end{equation}
where $\beta \in [0,1]$ is a weighting parameter that balances the contribution of spatial distance and copula-based similarity. To determine the optimal value of $\beta$, we run the PAM clustering for different possible values and evaluate clustering performance using silhouettes scores, that measure how well-defined the clusters are, and the Dunn index, that measure cluster separation. In addition, we select the optimal number of clusters using $k-$fold cross-validation.

\begin{algorithm}
\caption{Dissimilarity measure for the PAM Clustering Algorithm}\label{alg:fitting}
\begin{algorithmic}[1]
        \item Define each possible pair of regions $(i,j)$, for $i,j=1,\ldots,D$ and $i\neq j$. For each pair, estimate the parameters of the time-varying conditional copula model by optimizing the pseudo-likelihood function from Equation (\ref{eq:fulllik}).
        \item For each pair of census divisions $(i,j)$, 
        \begin{enumerate}
            \item Use the estimated model parameter to obtain the time-varying copula parameter $\hat{\theta}_t^{(i,j)}$, as well as the corresponding values of Kendall's tau, $\hat{\tau}_t^{(i,j)}$.
             \item Calculate the copula-based dissimilarity measure between each pair of census divisions as
        \begin{align*}
            d^{\text{copula}}_{(i,j),(k,l)} = \frac{1}{T} \sum_{t=1}^{T} \left| \hat{\tau}_t^{(i,j)} - \hat{\tau}_t^{(k,l)} \right|.
        \end{align*}
        \item Obtain the copula-dissimilarity measure at census division-level by taking the maximum dissimilarity
        \begin{align*}
             d_{i,j}^{\text{copula}} = \max_{(i,k)\in \mathcal{P}_i, (j,l)\in \mathcal{P}_j} d^{\text{copula}}_{(i,j),(k,l)}.
        \end{align*}
        \item Calculate the spatial distance-based dissimilarity measure using the Haversine formula:
        \begin{align*}
            d_{i,j}^{\text{spatial}} = R \arccos\big(\sin \psi_i \sin \psi_i + \cos \psi_i \cos \psi_j \cos(\lambda_i-\lambda_j)  \big)
        \end{align*}
        \item Define the combined dissimilarity measure as 
        \begin{align*}
            d^{\text{PAM}}_{(i,j)} = \beta d_{(i,j)}^{\text{spatial}} + (1-\beta) d^{\text{copula}}_{(i,j)}.
        \end{align*}
        \end{enumerate}
        \item Define a grid of $B$ candidate values $\beta_b$, with $b=1,\ldots,B$ such that $\beta_b \in [0,1]$, and select the number of clusters $K$ using a cross-validation method.
        \item For each candidate value $\beta_b$ in the grid, run the PAM clustering algorithm to define the $K$ clusters. Compute the the Dunn Index:
        \begin{align*}
            D(\beta_b) = \frac{\min\limits_{1 \leq k,m \leq K, k \neq m} \ \min\limits_{i \in C_k, j \in C_m} d_{i,j}}{\max\limits_{1 \leq k \leq K} \ \max\limits_{i,j \in C_k} d_{i,j}}
        \end{align*}
        \item Select the optimal value of $\beta$ that maximizes the Dunn Index:
        \begin{align*}
            \beta_{\text{opt}} = \arg \max_{\beta_b}D(\beta_b),
        \end{align*}
        such that the final combined dissimilarity measure is given by
        \begin{align*}
            d_{(i,j)} = \beta_{\text{opt}} d_{(i,j)}^{\text{spatial}} + (1-\beta_{\text{opt}}) d^{\text{copula}}_{(i,j)}.
        \end{align*}
        \item Run the PAM clustering algorithm using $d^{\text{PAM}}_{(i,j)}$ and $K$.
\end{algorithmic}
\end{algorithm}

\section{Application}
\label{sec:application}
We now apply the model described in Section \ref{sec:model} and the PAM clustering algorithm from Section \ref{sec:clustering} to the Ontario crop yields data from Section \ref{sec:data}. We begin this section by describing the forecasting algorithms used in combination with our model. We then present forecasts using five covariates, and end compare our results with forecasts using all 28 Core Climate Extreme Indices as covariates, as well as other forecast settings.

\subsection{Forecasting algorithm}
Before showing the results of our model on the Ontario data, we present the algorithms used to generate the crop yields forecasts. Algorithm \ref{alg:fitting} first describes how we fit the model and estimate all necessary parameters. We show the full forecasting procedure in Algorithm \ref{alg:forecast}.
\begin{algorithm}
\caption{Model fitting procedure}\label{alg:fitting}
\begin{algorithmic}[1]
        \item For $k$ in $1,...,K$ where $K$ is the total number of clusters used in the PAM clustering algorithm, fit Bayesian Structural Time Series for crop yields $Y_{k,t}$, where $t=1,\ldots,T$ are the time periods in the training data, and where $k$ represents the medoid of cluster $k$.
        \item Extract the fitted residuals from each model, $\epsilon^{(i)}_{k,t}$, with $i=1,\ldots,n-b$ where $n$ is the number of MCMC iterations performed to fit the BSTS, and $b$ the length of the burn-in period. 
        \item Transform each value of the residuals into pseudo-observations using the assumption that
        \begin{align*}
            \epsilon^{(i)}_{k,t}\sim\mathcal{N}(0,\sigma^{2}_{\epsilon^{(i)}_k}),
        \end{align*}
        where $\sigma^2_{\epsilon^{(i)}_k}$ are sampled from the posterior distribution derived with the BSTS. We have
        \begin{align*}
            u^{(i)}_{k,t} = F(\epsilon^{(i)}_{k,t}),
        \end{align*}
        where $F(.)$ is the cumulative distribution function of the Normal$(0,\sigma^{2}_{\epsilon^{(i)}_k})$ distribution.
        \item Keep only one value of the pseudo-observation per time $t$ by taking the average of the $n-b$ values derived for each time period in the previous step:
        \begin{align*}
            u_{k,t} = \frac{\sum_{i=1}^{n-b}u^{(i)}_{k,t}}{n-b}.
        \end{align*}
        \item Using the $T$ pseudo-observations obtained in the previous step, optimize the pseudo-likelihood function given in Equation (\ref{eq:fulllik}) to obtain all necessary parameter estimates.
\end{algorithmic}
\end{algorithm}

\begin{algorithm}
\caption{Forecasting procedure}\label{alg:forecast}
\begin{algorithmic}[1]
        \item Fit the model and retrieve all parameter estimates using Algorithm \ref{alg:fitting}.
        \item For forecasting periods $t^*=T+1,\ldots,T^*$, where $T$ is the total number of periods in the training data,
        \begin{enumerate}
            \item For risk factor $m=1,\ldots,M$, 
            \begin{itemize}
                \item simulate a new observation $\epsilon_{d,m,t^*}\sim\mathcal{N}(0,\hat{\sigma}_{\mu_{d,m}})$
                \item use $\epsilon_{d,m,t^*}$ to calculate a new value of the GEV location parameter:
                \begin{align*}
                    \mu_{d,m,t^*} = \hat{\phi}_{d,m}\mu_{d,m,t^*-1} + \epsilon_{d,m,t^*}
                \end{align*}
                \item simulate a new realisation of the covariate:
                \begin{align*}
                    z_{d,m,t^*} = F_{Z{m,t}}^{-1}(u_{d,m};\mu_{d,m,t^*},\hat{\sigma}_{d,m},\hat{\xi}_{d,m}), \hspace{0.3cm}\text{with $u_{d,m}\sim\mathcal{U}(0,1)$}.
                \end{align*}
            \end{itemize}
            \item Retrive the new realisation of $X_{m,t^*}$ using
            \begin{align*}
                x_{m,t^*} = \max_{d=1,\ldots,D}z_{d,m,t^*}.
            \end{align*}
            \item Compute the new value of the copula dependence parameter:
            \begin{align*}
                \theta_{t^*}(\vect{x}_{t^*}) = g\Big(\hat{\omega} + \hat{\alpha}\theta_{t^*-1} + \sum_{m=1}^{M}\hat{\gamma}_m x_{m,t^*} \Big).
            \end{align*}
            \item For $d=1,\ldots,D$, simulate a $1\times(n-b)$ vector of pseudo-observations $u^{(i)}_{d,t^*}$ from the copula with dependence parameter $\theta_{t^*}(\vect{x}_{t^*})$.
            \item For each $i=1,\ldots,n-b$, transform $u^{(i)}_{d,t^*}$ back into residuals using 
            \begin{align*}
                \hat{\epsilon}^{(i)}_{d,t^*} = F^{-1}(u^{(i)}_{d,t^*}),
            \end{align*}
            where $F^{-1}(.)$ is the inverse cumulative distribution function of the Normal$(0,\sigma^{2}_{\epsilon^{(i)}_d})$ distribution.
            \item Combine outputs from the BSTS models with the residuals calculated in the previous step to obtain the new crop yield forecast in region $d$:
            \begin{align*}
                \hat{y}_{d,t^*} = l_{d,t^*} + s_{d,t^*} + \sum_{i=1}^{p}\hat{\psi}_{d,l}e_{d,t^*-l} + \sum_{m=1}^{M}\hat{\beta}_{d,m,t^*}z_{d,m,t^*} + \hat{\epsilon}_{d,t^*},
            \end{align*}
            where $\hat{\epsilon}_{d,t^*}$ is sampled from the $n-b$ simulated values $\hat{\epsilon}^{(i)}_{d,t^*}$.
        \end{enumerate}
        \item Repeat step 2 a sufficiently large enough amount of times to obtain a distribution of the forecasted yields in all $D$ regions for time periods $t^*=T+1,\ldots,T^*$.
\end{algorithmic}
\end{algorithm}

\subsection{Forecasts}
We first consider the five covariates selected and described in Section \ref{sec:data}, namely the annual maximum number of consecutive days with precipitations below 1 mm ($X_1$), the number of days with precipitations above 10 mm ($X_2$), the number of frost days ($X_3$), the annual mean of daily mean temperatures ($X_4$), and the number of days with temperatures exceeding 25$^{\circ}C$ ($X_5$).

We begin by fitting Bayesian Structural Time Series (BSTS) models to the crop yields of the 24 census divisions considered in this study. For all models, we select the commonly chosen prior uniform distribution $\mathcal{U}(-1,1)$ for the coefficients of the auto-regressive component $\psi_d$, with $d=1,\ldots,24$. We select the Inverse-Gamma distributions for the variance parameters, i.e. $\sigma^2_{\epsilon_d}$, $\sigma^2_{\nu_d}$, $\sigma^2_{\zeta_d}$, $\sigma^2_{\eta_d}$ and $\sigma^2_{\lambda_d}$. We retrieve the posterior distributions for all these estimators by performing 10,000 MCMC simulations. Note that, since we work with yearly crop yield data, we do not include any seasonal component in the BSTS models. 

We then extract the residuals $u_{d,t}$ from these $D=24$ marginal models. We use them to run the PAM clustering algorithm, systematically evaluating every possible pair of census divisions and comparing them based on both their spatial dissimilarity (using the Haversine formula) and our novel copula-based dissimilarity measure. The optimal number of clusters is determined to be two, with medoids identified in the census divisions of Dufferin and Wellington (Prince Edward). The resulting clusters are displayed on the map in Figure \ref{fig:map}, with medoids marked by stars. Cluster 1, centered around Dufferin, is shown in red, while Cluster 2, surrounding Wellington, is depicted in blue.

These results are not only statistically significant but also geographically meaningful, as they align closely with the agricultural profile of Ontario. According to the Ontario Ministry of Agriculture, Food and Rural Affairs (OMAFRA), Ontario's agriculture is primarily concentrated in two distinct regions: the fertile farmland of Southwestern Ontario, considered prime agricultural area, including the Greater Toronto Area (GTA) and westward, and the less agriculturally intensive region to the east of Toronto (\cite{OMAFRA2023}). The red cluster, which encompasses the Dufferin area and extends west of Toronto, correlates with Southwestern Ontario's highly productive agricultural belt, known for its rich soils and high crop yields. This region supports significant agricultural activities, including the production of grains, oilseeds, and horticultural crops. Conversely, the blue cluster, located east of Toronto around Wellington, aligns with a region characterized by lower agricultural density and more diverse land use, including forests and residential areas.

The fact that the PAM algorithm successfully identified these two distinct agricultural zones demonstrates its ability to capture complex spatial and dependency structures within the crop yield data. By integrating spatial information with copula-based measures of dependence, our method reflects the real-world geographical segmentation of agricultural production in Ontario, highlighting the algorithm's robustness and interpretative value in agricultural modeling.

\begin{figure}[h!]
\centering
\includegraphics[width=.6\textwidth]{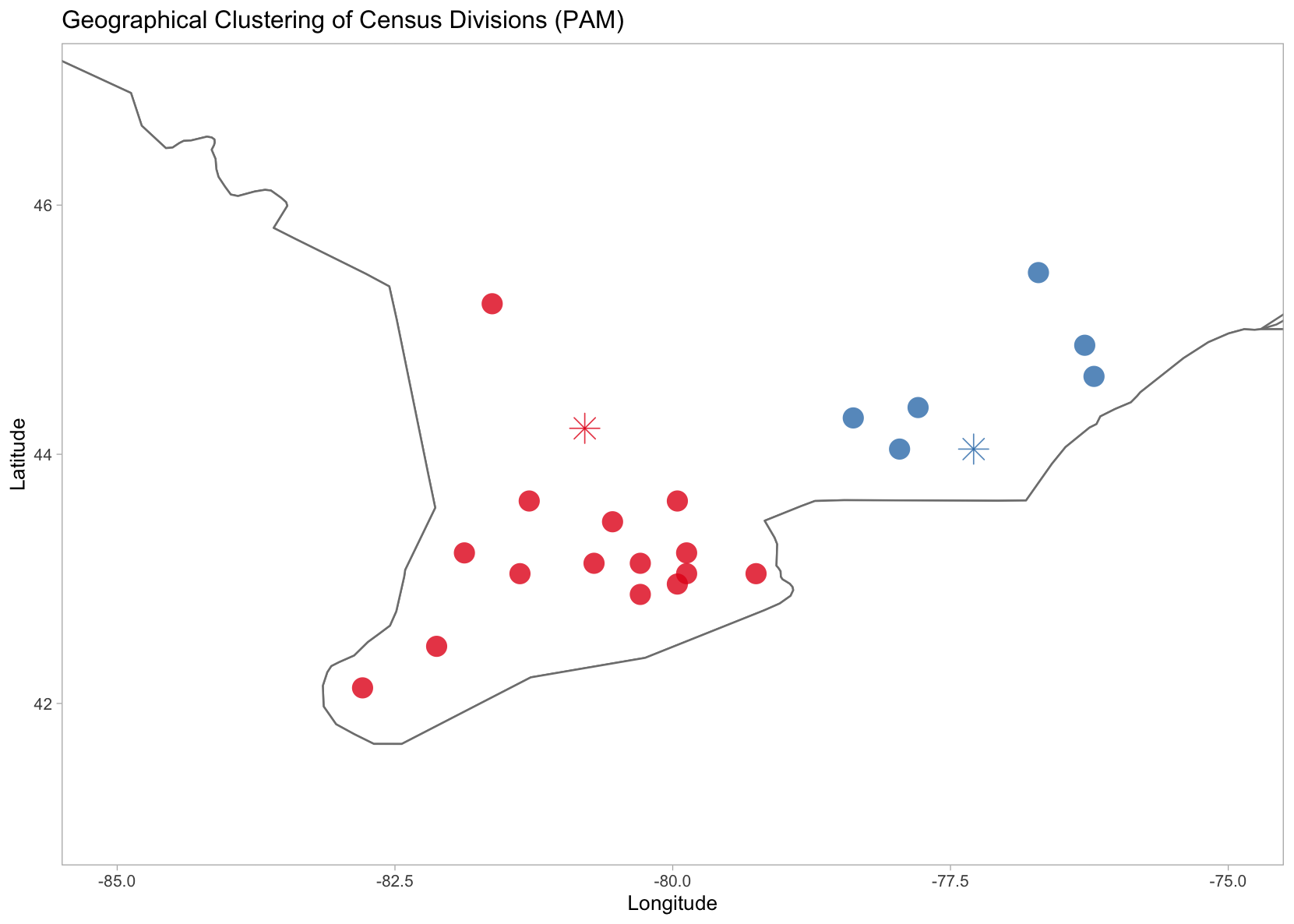}
\caption{Clusters formed by the PAM algorithm, shown on a map of Ontario.}
\label{fig:map}
\end{figure}

To evaluate the added value of incorporating dependence structures into the clustering process via the copula-based dissimilarity measure, we compared these results with those obtained using a more traditional spatial-based only dissimilarity measure. Interestingly, both methods produced the same two clusters, suggesting that spatial proximity remains a dominant factor in shaping agricultural similarity across regions. However, we observe a difference in the medoid selected for the first cluster. While the combined copula and spatial dissimilarity approach identifies Dufferin as the medoid, the spatial-only method selects Brantford. This shift is intuitive, as Brantford is more centrally located within the cluster from a purely geographical perspective, reflecting the fact that spatial proximity is the sole driver of similarity in the spatial-only approach.

\begin{figure}[h!]
\centering
\includegraphics[width=.5\textwidth]{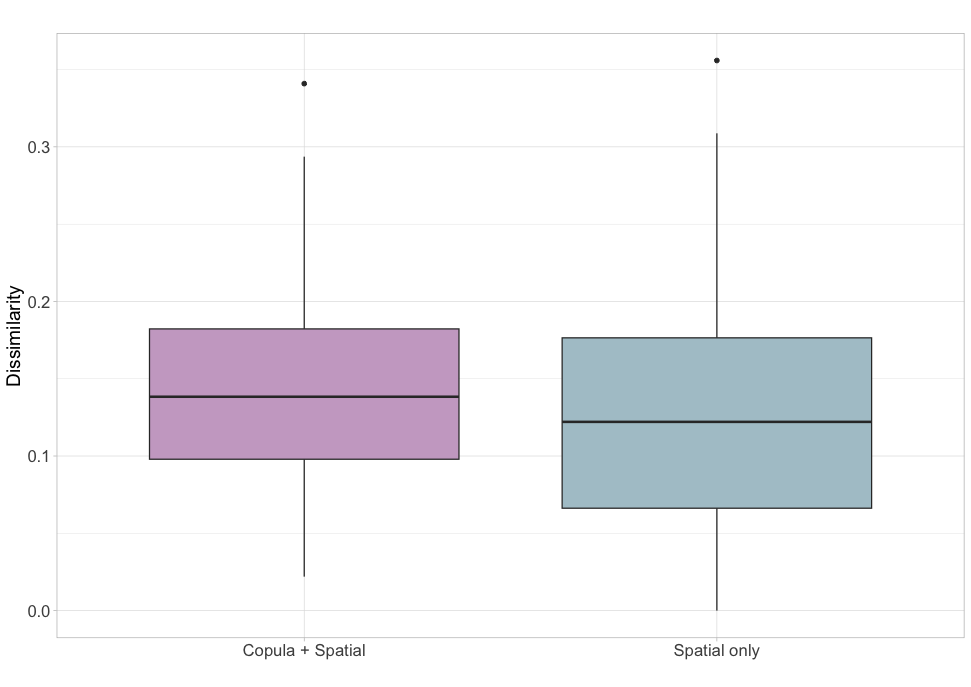}
\caption{Boxplots of the within-cluster dissimilarities.}
\label{fig:boxplot}
\end{figure}

Figure \ref{fig:boxplot} displays the boxplots of within-cluster dissimilarities obtained with both approaches. We observe that the clusters produced using our combined dissimilarity measure are more cohesive. Specifically, the distribution of within-cluster distances under our approach shows a reduced variability compared to the spatial-only method. This indicates that, even though the cluster assignments remain stable, our copula-based dissimilarity matrix leads to tighter and more internally consistent clusters. It confirms that the inclusion of dependence dynamics, captured through time-varying copulas, adds meaningful refinement to the clustering process by ensuring that regions grouped together are not only geographically close but also exhibit similar patterns in crop yield dependencies over time.

With our two clusters, we then optimize the pseudo-likelihood function from Equation (\ref{eq:fulllik}), for $D=K=2$ and $M=5$. The conditional copula parameter is given by
\begin{align*}
    \theta_t(\vect{X}_t) = g(\omega + \alpha \theta_{t-1}+\gamma_1X_{1,t}+\gamma_2X_{2,t}+\gamma_3X_{3,t}+\gamma_4X_{4,t}+\gamma_5X_{5,t}),
\end{align*}
while the dynamic GEV density functions for $Z_{d,m,t}$, with $k=1,\ldots,5$ are given by Equation (\ref{eq:GEV}), for $d=1,2$.

We can then optimize the pseudo-likelihood function from Equation (\ref{eq:fulllik}). We select the initial parameter values using the method of moment matching, ensuring that the starting estimates are consistent with the empirical moments of the data. We use five different copulas to model the dependence between our crop yields: one elliptical copula, namely the Gaussian copula, and four Archimedean copulas, the Clayton, Frank, Gumbel and Joe copulas. Table \ref{tab:model_results} presents the average mean squared error (AMSE) from each model, computed as
\begin{align}
\label{eq:AMSE}
    \text{AMSE} = \frac{1}{2n} \sum_{i=1}^{n} \sum_{d=1}^2(\hat{y}_{d,i} - y_{d,i})^2.
\end{align}

\begin{table}[ht]
\centering
\begin{tabular}{lc}
\toprule
Copula &  AMSE \\ \midrule
Gaussian & 399.2726\\
Clayton & 378.8733\\
Frank & 365.8299\\
Gumbel & 364.8268\\
Joe & 372.0145\\
\bottomrule
\end{tabular}
\caption{Model comparison}
\label{tab:model_results}
\end{table}

The results from Table \ref{tab:model_results} indicate that the Gumbel copula provides the best fit for modeling the dependence structure in the data, as it minimizes the Average Mean Squared Error (AMSE). This suggest that the dependence is characterized by upper tail and lower tail dependence, where extreme high and low values in the marginal distributions are more likely to occur simulateneously. The Gaussian copula shows the highest AMSE of 399.2726, suggesting that the assumption of linear dependence might not fully capture the complex dependencies present in the data. This reinforces the need for more flexible dependency structures in agricultural yield forecasting, especially under the influence of extreme climatic conditions.

\begin{figure}[h!]
\centering
\includegraphics[width=1\textwidth]{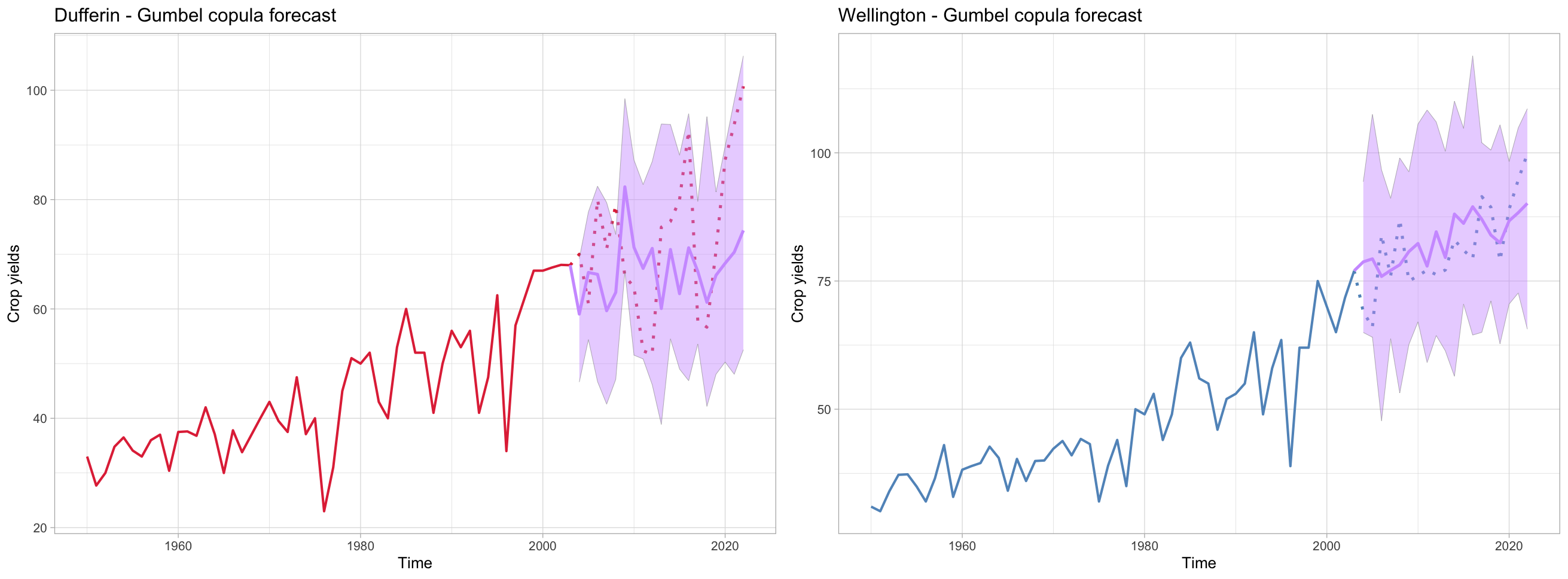}
\caption{Crop yields simulations for Dufferin (left) and Wellington (right).}
\label{fig:sim}
\end{figure}

Figure \ref{fig:sim} shows the results of 1,000 simulations for the crop yields of both census divisions from 2004 to 2022, using the conditional time-varying Gumbel copula. In both plots, the purple line represents the average of the simulations, the shaded area shows their $95\%$ confidence interval and the dotted lines stand for the actual crop yields observed in the validation set. 

The forecasted crop yields for both Dufferin and Wellington show a strong alignment with the observed data, demonstrating the effectiveness of the Gumbel copula model in capturing the dependence structure between census divisions. The forecasts not only track the general upward trend in crop yields but also adapt well to short-term fluctuations.

The confidence intervals, represented by the shaded purple areas, provide a robust quantification of uncertainty. They successfully encompass the full range of observed values, including both the extreme peaks and troughs in crop yields. This indicates that the model is appropriately capturing the inherent volatility in agricultural production, which can be heavily influenced by unpredictable climatic conditions.

A noteworthy observation is the broader confidence interval for the Wellington region compared to Dufferin. This suggests a higher uncertainty in crop yield predictions for Wellington, potentially reflecting greater sensitivity to external risk factors such as climate variability. This difference in uncertainty levels between the two regions aligns with the known agricultural and environmental characteristics of Ontario. The Dufferin region, situated in a more agriculturally intensive and stable area, exhibits less volatility, whereas Wellington, located towards the less productive eastern zone, demonstrates more variability in crop yields.

\subsubsection{Impact of dependence modeling on crop yield forecasts}
\begin{figure}[h!]
\centering
\includegraphics[width=1\textwidth]{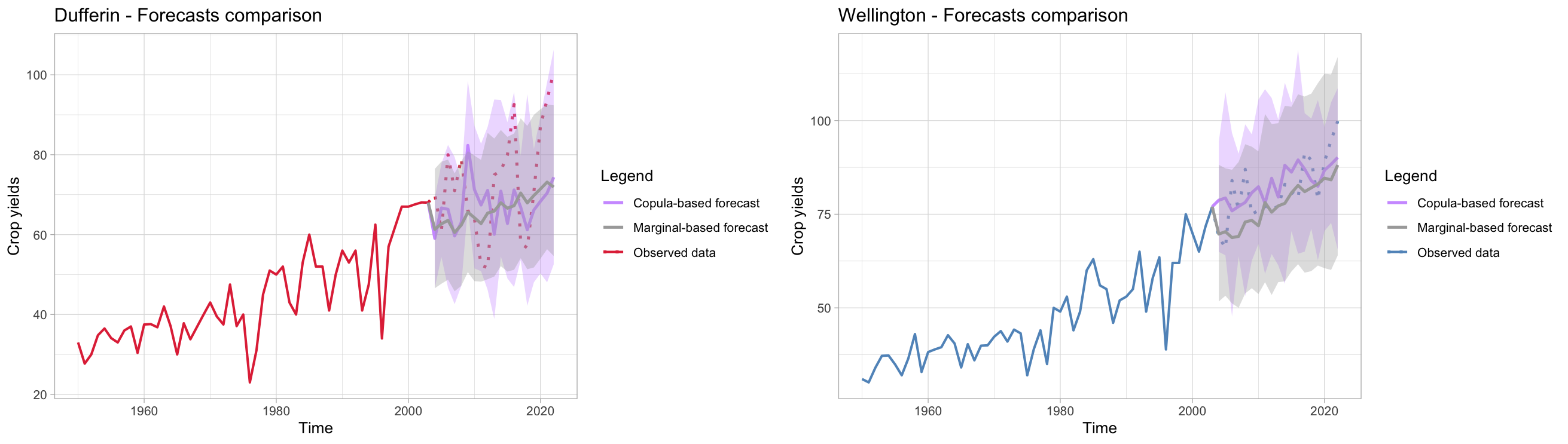}
\caption{Comparison of forecasts using the copula-based model (purple) and independence (grey).}
\label{fig:INDCOMP}
\end{figure}

We now compare the forecasts obtained with our time-varying conditional copula model to an approach that assumes independence betwen the regions. More specifically, we replace the Gumbel copula from the previous section by the independence copula. The comparison between the forecasts resulting from both approaches is presented in Figure \ref{fig:INDCOMP}.

The comparison between the copula-based forecasts (in purple) and the marginal-based forecasts (in grey) provides valuable insights into the impact of modeling both spatial and temporal dependence through the time-varying conditional copula approach. For both Dufferin and Wellington, the copula-based forecasts generally follow the observed crop yields more closely than the marginal-based forecasts, highlighting the added predictive power of accounting for evolving dependencies over time and between regions.

The confidence intervals associated with the copula-based forecasts are broader, reflecting the model’s ability to incorporate the uncertainty related to the dependence structure both across census divisions and over time. This is particularly evident in the Dufferin region, where the increased variation in crop yields is well captured by the purple shaded area. In contrast, the marginal-based model, which assumes spatial independence, produces narrower confidence intervals that may underestimate the true variability of crop yields, especially in the presence of correlated extreme events.

Another key observation is the difference in the forecasted trajectories. While the marginal-based forecasts tend to smooth out short-term variations and provide a more linear projection, the copula-based forecasts exhibit greater flexibility, adapting dynamically to changes in crop yields influenced by shared climate risks and temporal trends. This adaptability is particularly beneficial for forecasting scenarios where extreme climatic conditions may create sudden shifts in agricultural output.

Overall, these plots demonstrate the advantage of using a copula-based approach to better capture both the trends and the uncertainty in crop yields. By modeling the evolving spatial and temporal dependencies, the copula-based model enhances the robustness of the forecasts, offering more reliable predictions that could support agricultural risk management and decision-making processes.

\subsubsection{Impact of extreme event modeling on crop yield forecasts}

\begin{figure}[h!]
\centering
\includegraphics[width=1\textwidth]{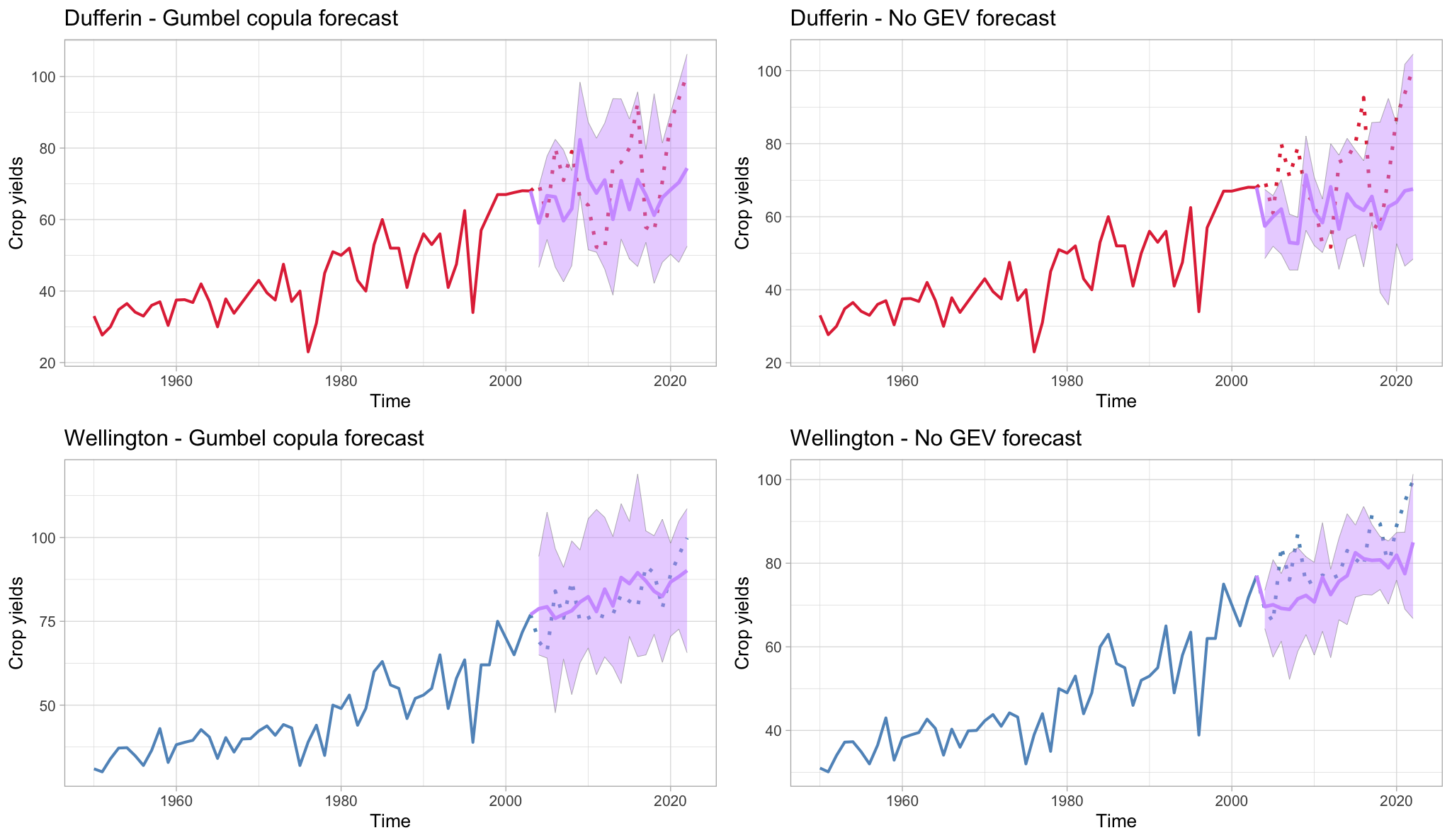}
\caption{Comparison of forecasts using the copula-based model with (left) and without (right) dynamic GEV for the covariates.}
\label{fig:GEVCOMP}
\end{figure}

In Figure \ref{fig:GEVCOMP}, we compare our approach of forecasting crop yields using the time-varying conditional copula model where the climatic covariates are modeled using dynamic Generalized Extreme Value (GEV) distributions (left panels) to an approach where the covariates are directly taken as observations without a GEV model (right panels). The dynamic GEV model allows for a more flexible representation of extreme climatic events by modeling the covariates' behavior over time, especially their tails and the influence of extreme values. In contrast, using the raw covariate observations without modeling assumes that the covariate values are perfectly known and do not account for the uncertainty in their distribution, particularly under extreme scenarios. This comparison highlights the effect of explicitly modeling extreme weather and climatic events within our forecasting framework.

The first observation from the plots is that the forecasts with the GEV models (left panels) provide wider confidence intervals compared to the forecasts without the GEV models (right panels). This broader uncertainty band is indicative of the model capturing not only the expected climatic influences but also the potential variability introduced by extreme weather events. This is particularly relevant for agricultural forecasting, where extreme events like droughts or floods can lead to significant deviations in crop yields.

For the Dufferin region, the GEV-based model (top-left panel) achieves a better alignment with the observed data, especially during periods of high variability. The confidence intervals encompass most of the observed peaks and troughs in the data, suggesting that the model effectively accounts for the underlying uncertainty and the impact of extreme climatic conditions. The model without GEV (top-right panel), however, generates narrower confidence intervals that often miss the extreme variations, potentially underestimating the risk of extreme outcomes in crop yields.

In the Wellington region, similar patterns are observed. The GEV-based model (bottom-left panel) shows a more flexible adaptation to observed yield changes, with confidence intervals that appropriately reflect the variability in the data. In contrast, the non-GEV model (bottom-right panel) tends to provide a smoother forecast with narrower confidence bounds, which may overlook critical variations caused by extreme weather events. For both census divisions, the non-GEV forecasts also seem to underestimate the observed data on average compared to the GEV forecats. 

Overall, incorporating dynamic GEV models for the covariates within the time-varying conditional copula framework provides a more robust and cautious forecasting approach. It captures not only the dependence structure across regions and time but also the potential impacts of extreme climatic conditions.

\subsubsection{Results summary}
To summarise the results from the three different forecasting approaches considered in this section, we calculate for each census division the average mean squared error (AMSE) and the average mean absolute errors (AMAE) of the forecasts. More specifically, for census divisions $d$ with $d=1,2$, we use
\begin{align*}
    \text{AMSE}_d &= \frac{1}{N} \sum_{i=1}^{N}\Bigg( \frac{1}{T}\sum_{t=1}^{T} (y_{d,t} - \hat{y}^{(n)}_{d,t})^2\Bigg) \\
    \text{AMAE}_d &= \frac{1}{N}\sum_{n=1}^{N}\Bigg(\frac{1}{T}\sum_{t=1}^{T} \vert y_{d,t} - \hat{y}_{d,t}^{(n)} \vert\Bigg),
\end{align*}
where $N$ represents the total number of forecasts simulations performed and $T$ is the number of forecasted periods. We present the results for Dufferin and Wellington in Table \ref{tab:AMAE}.

\begin{table}[ht]
\centering
\begin{tabular}{lcccc}
\toprule
\multirow{2}{*}{Model}  &  \multicolumn{2}{c}{AMAE} & \multicolumn{2}{c}{AMSE} \\
& Dufferin & Wellington & Dufferin & Wellington \\ \midrule
Gumbel copula with GEV & 11.0088 & 8.0546 & 163.6467 & 94.4161\\
Independence copula with GEV & 11.6379 & 13.5008 & 187.9113 & 220.5978\\
Gumbel copula without GEV & 11.8584 & 16.7879 & 190.1951 & 336.6769\\
\bottomrule
\end{tabular}
\caption{Average mean absolute errors (AMAE) and average mean squared error (AMSE) of different models for both census divisions.}
\label{tab:AMAE}
\end{table}

Our model combining a Gumbel copula with a Generalized Extreme Value (GEV) framework achieves a strong performance overall, particularly in Wellington, where it records the lowest AMAE of 8.0546 and lowest AMSE of 94.4161. This indicates that modeling both the dependence structure between regions through a time-varying conditional copula and accounting for extreme climatic events with a GEV distribution contribute significantly to forecast accuracy. Similar results are observed for the Dufferin census division, although the difference in performance between the models is less marked. 

The Independence copula with GEV performs worse than the Gumbel copula with GEV, especially in Wellington, where the AMAE is significantly higher. This suggests that assuming independence between crop yields in different regions is a poor assumption and leads to less accurate predictions.

The Gumbel copula without GEV is the worst-performing model overall. The AMAE values are the highest for both Dufferin (11.8584) and Wellington (16.7879), and the AMSE values are also the worst (190.1951 for Dufferin and 336.6769 for Wellington). This demonstrates that failing to model extreme climate effects properly leads to larger forecast errors, especially in squared error terms. 

Since the Gumbel copula without GEV has the highest AMSE, this suggests that it occasionally produces extremely poor forecasts, which drastically increases the squared error. The Independence copula with GEV also exhibits high AMSE values, indicating that assuming independence between crop yields can lead to extreme mispredictions when strong dependencies exist.

\color{black}
\section{Conclusion}
\label{sec:conclusion}
We present a novel forecasting framework for crop yields that accounts for both spatio-temporal dependence and the effects of extreme weather events and climate change. Specifically, our approach integrates a Bayesian methodology for modeling the marginal distributions of crop yields with a pseudo-likelihood optimization framework for the joint dependence structure. Marginal crop yields are modeled using Bayesian Structural Time Series (BSTS), which provide flexibility in capturing trends and covariate effects, perform well in small-sample settings, and offer robust uncertainty quantification. The dependence structure is then modeled using a time-varying conditional copula, where the copula parameter dynamically evolves as a function of its past values and extreme weather covariates. To further improve the representation of extreme climatic effects, we model the covariates using dynamic Generalized Extreme Value (GEV) distributions, allowing the framework to incorporate evolving risk factors more effectively.

A key contribution of this work is the development of a new dissimilarity measure for clustering, which significantly improves the scalability of our model. We introduce a combined dissimilarity measure, which integrates both spatial and copula-based dependence metrics, offering a more informative and flexible criterion for clustering agricultural regions. Using Partitioning Around Medoids (PAM) clustering, we demonstrate the efficiency of this approach by identifying the two most distinct and influential agricultural regions in Ontario, allowing us to reduce the dimensionality of our dependence structure while retaining the most critical spatial dependencies. This not only enhances computational efficiency but also facilitates the application of our model on a larger scale, making it suitable for forecasting across broader agricultural landscapes.

We apply this framework to winter wheat crop yield data from multiple census divisions in Ontario, incorporating 28 core climate extreme indices developed by the ETCCDI as weather covariates. Our analysis evaluates four Archimedean copulas and one elliptical copula. The results indicate that Archimedean copulas provide a better fit than the Gaussian copula, underscoring the importance of capturing tail dependencies in crop yield data.

To assess the performance of our approach, we compare the copula-based forecasts with those obtained solely from BSTS models applied to the marginal distributions, without incorporating dependence structures or modeling covariates as dynamic GEV distributions. The results confirm that copula-based forecasts exhibit narrower confidence intervals, indicating a reduction in forecast uncertainty. This improvement is likely due to the explicit modeling of spatial dependencies and the dynamic evolution of the copula parameter, as well as the enhanced representation of climate risks through dynamic GEV modeling. However, the magnitude of the improvement varies across regions, suggesting that further refinements—such as allowing for more flexible dependence structures—could enhance predictive accuracy.

The comparative evaluation of forecasting errors further reinforces the effectiveness of our approach. The Gumbel copula with GEV consistently outperforms the alternative models, achieving the lowest AMAE and AMSE values across all regions. This result highlights the importance of accounting for both spatial dependence and extreme weather effects, as ignoring these components leads to higher forecast errors, particularly in squared error terms. The independence copula with GEV performs worse than the Gumbel copula, emphasizing that assuming independence between regions introduces significant forecast bias. Similarly, the Gumbel copula without GEV exhibits the highest AMSE values, suggesting that failing to model extreme weather leads to large, infrequent errors that disproportionately affect squared loss metrics. These findings confirm that jointly modeling spatio-temporal dependencies, extreme weather risks, and clustering-based dimensionality reduction enhances forecasting accuracy and reliability.

Future work could explore refinements to the dependence structure by moving beyond a single Archimedean copula when working in higher dimensions. Since Archimedean copulas are governed by a single dependence parameter, they may oversimplify complex dependence patterns across multiple regions, particularly those that are geographically distant. An alternative could involve using vine copulas, which allow for pairwise dependence modeling while still capturing tail dependencies—an important feature in agricultural forecasting. Additionally, further work could incorporate additional covariates, test the model on different crops or regions, and refine clustering techniques to optimize computational efficiency in large-scale applications.

Overall, this study introduces a comprehensive, interpretable, and scalable forecasting framework, integrating Bayesian time series modeling, time-varying conditional copulas, dynamic extreme value modeling, and a clustering methodology. By implementing these methods in an open-source R package, we provide researchers and policymakers with a reproducible and adaptable tool for improving crop yield forecasts under evolving climate risks. These advancements contribute to more accurate and resilient agricultural forecasting, with direct applications in food security planning, agricultural risk management, and climate adaptation strategies.

\vspace{1cm}
\noindent\textit{The R package implementing our proposed methodology is publicly available for download on CRAN, ensuring easy reproducibility and application of our framework.}
\url{https://cran.r-project.org/web/packages/STCCGEV/index.html}

\clearpage
\bibliography{main}

\begin{thebibliography}{34}
\providecommand{\natexlab}[1]{#1}
\providecommand{\url}[1]{\texttt{#1}}
\expandafter\ifx\csname urlstyle\endcsname\relax
  \providecommand{\doi}[1]{doi: #1}\else
  \providecommand{\doi}{doi: \begingroup \urlstyle{rm}\Url}\fi

\bibitem[Alidoost et~al.(2019)Alidoost, Su, and Stein]{alidoost2019}
F.~Alidoost, W.~Su, and A.~Stein.
\newblock Evaluating the effects of climate extremes on crop yield, production and price using multivariate distributions: A new copula application.
\newblock \emph{Weather and Climate Extremes}, 26, 2019.

\bibitem[Boulin et~al.(2025)Boulin, Di~Bernardino, Laloë, and Toulemonde]{boulin2025}
A.~Boulin, E.~Di~Bernardino, T.~Laloë, and G.~Toulemonde.
\newblock Identifying regions of concomitant compound precipitation and wind speed extremes over europe.
\newblock \emph{Journal of the Royal Statistical Society Series C: Applied Statistics}, 2025.

\bibitem[Boyer et~al.(2014)Boyer, Brorsen, and Tumuslime]{boyer2014}
C.~Boyer, B.~Brorsen, and E.~Tumuslime.
\newblock Modeling skewness with the linear stochastic plateau model to determine optimal nitrogen rates.
\newblock \emph{Agricultural Economics}, 46\penalty0 (1):\penalty0 1--10, 2014.

\bibitem[Celis et~al.(2024)Celis, Xiao, Wagle, Adler, and White]{Celis2024}
J.~Celis, X.~Xiao, P.~Wagle, P.~Adler, and P.~White.
\newblock A review of yield forecasting techniques and their impact on sustainable agriculture.
\newblock In \emph{Sustainable Agriculture: Emerging Trends and Innovations}, pages 139--164. Springer, 2024.

\bibitem[Chemere et~al.(2018)Chemere, Kim, Peng, Kim, and Sung]{chemere2018}
B.~Chemere, M.~Kim, J.~Peng, B.-Y. Kim, and K.-C. Sung.
\newblock Detecting dry matter yield trend of whole crop maize considering the climatic factors in the republic of korea.
\newblock \emph{Grassland Science}, 65\penalty0 (2):\penalty0 116--124, 2018.
\newblock \doi{10.1111/grs.12220}.

\bibitem[Cheung et~al.(2024)]{cheung2024}
Y.~Cheung et~al.
\newblock Analyzing the impacts of extreme weather factors using the actuaries climate index: A bonus–malus approach for crop insurance.
\newblock \emph{Actuarial Studies and Climate Risk Journal}, 2024.

\bibitem[Coles(2001)]{Coles2001}
S.~Coles.
\newblock \emph{An Introduction to Statistical Modeling of Extreme Values}.
\newblock Springer, London, 2001.

\bibitem[De~Paola et~al.(2018)De~Paola, Giugni, Pugliese, Annis, and Nardi]{depaola2018}
F.~De~Paola, M.~Giugni, F.~Pugliese, A.~Annis, and F.~Nardi.
\newblock Gev parameter estimation and stationary vs. non-stationary analysis of extreme rainfall in african test cities.
\newblock \emph{Hydrology}, 5\penalty0 (2):\penalty0 28, 2018.

\bibitem[Feng et~al.(2023)Feng, Tian, Cong, and Zhao]{feng2023method}
X.~Feng, H.~Tian, J.~Cong, and C.~Zhao.
\newblock A method review of the climate change impact on crop yield.
\newblock \emph{Frontiers in Forests and Global Change}, 6:\penalty0 1198186, 2023.

\bibitem[Fischer and Knutti(2015)]{fischer2015}
E.~Fischer and R.~Knutti.
\newblock Anthropogenic contribution to global occurrence of heavy-precipitation and high-temperature extremes.
\newblock \emph{Nature Climate Change}, 5\penalty0 (6):\penalty0 560--564, 2015.

\bibitem[Gilleland and Katz(2016)]{gilleland2016}
E.~Gilleland and R.~W. Katz.
\newblock Extending extreme value theory with application to climate change.
\newblock \emph{Weather and Climate Extremes}, 13:\penalty0 1--9, 2016.

\bibitem[{Government of Ontario}(2025)]{OMAFRA2023}
{Government of Ontario}.
\newblock Prime agricultural areas, 2025.
\newblock URL \url{https://www.ontario.ca/page/prime-agricultural-areas}.
\newblock Accessed: 2025-02-13.

\bibitem[Gumbel(1958)]{Gumbel1958}
E.~J. Gumbel.
\newblock \emph{Statistics of Extremes}.
\newblock Columbia University Press, New York, 1958.

\bibitem[Gunawan and Gunardi(2023)]{Katarina2023}
K.~Gunawan and G.~Gunardi.
\newblock Optimising bayesian structural time series for stock price forecasting.
\newblock In \emph{Proceedings of the 5th International Conference on Intelligent and Advanced Systems (ICIAS)}, volume 987 of \emph{Lecture Notes in Electrical Engineering}, pages 273--281. Springer, 2023.

\bibitem[Huerta and Sansó(2007)]{huerta2007}
G.~Huerta and B.~Sansó.
\newblock Time-varying models for extreme values.
\newblock \emph{Environmental and Ecological Statistics}, 14:\penalty0 285--299, 2007.

\bibitem[Kim et~al.(2019)Kim, Chemere, and Sung]{kim2019}
M.~Kim, B.~Chemere, and K.-C. Sung.
\newblock Effect of heavy rainfall events on the dry matter yield trend of whole crop maize (zea mays l.).
\newblock \emph{Agriculture}, 9\penalty0 (4):\penalty0 75, 2019.
\newblock \doi{10.3390/agriculture9040075}.

\bibitem[Liu et~al.(2021)Liu, Yu, and Liu]{Liu2021}
Y.~Liu, J.~Yu, and J.~Liu.
\newblock Bayesian structural time series model for biomedical sensor data analysis.
\newblock \emph{PLOS Computational Biology}, 17\penalty0 (9):\penalty0 e1009303, 2021.

\bibitem[Lobell et~al.(2011)Lobell, Schlenker, and Costa-Roberts]{lobell2011}
D.~B. Lobell, W.~Schlenker, and J.~Costa-Roberts.
\newblock Climate trends and global crop production since 1980.
\newblock \emph{Science}, 333\penalty0 (6042):\penalty0 616--620, 2011.

\bibitem[Nasri and R{\'e}millard(2019)]{nasri2019copula}
B.~R. Nasri and B.~N. R{\'e}millard.
\newblock Copula-based dynamic models for multivariate time series.
\newblock \emph{Journal of Multivariate Analysis}, 172:\penalty0 107--121, 2019.

\bibitem[Nelsen(2006)]{nelsen2006}
R.~Nelsen.
\newblock \emph{An Introduction to Copulas}.
\newblock Springer, New York, 2 edition, 2006.

\bibitem[Palacios~Rodriguez et~al.(2023)Palacios~Rodriguez, Di~Bernardino, and Mailhot]{palacios2023}
F.~Palacios~Rodriguez, E.~Di~Bernardino, and M.~Mailhot.
\newblock Smooth copula-based generalized extreme value model and spatial interpolation for extreme rainfall in central eastern canada.
\newblock \emph{Environmetrics}, 34\penalty0 (3), 2023.

\bibitem[Patton(2006)]{patton2006}
A.~Patton.
\newblock Copula-based models for financial time series.
\newblock In \emph{Mikosch, T., Kreiß, JP., Davis, R., Andersen, T. (eds) Handbook of Financial Time Series}, Berlin, Heidelberg, 2006. Springer.

\bibitem[Ray et~al.(2015)Ray, Gerber, MacDonald, and West]{Ray2015}
D.~K. Ray, J.~S. Gerber, G.~K. MacDonald, and P.~C. West.
\newblock Climate variation explains a third of global crop yield variability.
\newblock \emph{Nature Communications}, 6\penalty0 (1):\penalty0 5989, 2015.

\bibitem[Reddy and Sureshbabu(2020)]{reddy2020}
P.~C.~S. Reddy and A.~Sureshbabu.
\newblock An applied time series forecasting model for yield prediction of agricultural crop.
\newblock In V.~Reddy, V.~Prasad, J.~Wang, and K.~Reddy, editors, \emph{Soft Computing and Signal Processing. ICSCSP 2019}, volume 1118 of \emph{Advances in Intelligent Systems and Computing}, pages 163--171. Springer, 2020.

\bibitem[Rosenzweig et~al.(2014)]{rosenzweig2014}
C.~Rosenzweig et~al.
\newblock Assessing agricultural risks of climate change in the 21st century.
\newblock \emph{Proceedings of the National Academy of Sciences}, 111\penalty0 (9):\penalty0 3268--3273, 2014.

\bibitem[Salvadori et~al.(2014)]{salvadori2014}
G.~Salvadori et~al.
\newblock A multivariate copula-based framework for dealing with hazard scenarios and failure probabilities.
\newblock \emph{Water Resources Research}, 50\penalty0 (4):\penalty0 3900--3921, 2014.

\bibitem[Schlenker and Roberts(2009)]{schlenker2009}
W.~Schlenker and M.~Roberts.
\newblock Nonlinear temperature effects indicate severe damages to u.s. crop yields under climate change.
\newblock \emph{Proceedings of the National Academy of Sciences}, 106\penalty0 (37):\penalty0 15594--15598, 2009.

\bibitem[Scott and Varian(2014)]{scott2014}
S.~L. Scott and H.~R. Varian.
\newblock Predicting the present with bayesian structural time series.
\newblock \emph{International Journal of Mathematical Modelling and Numerical Optimisation}, 5\penalty0 (12):\penalty0 4--23, 2014.

\bibitem[Shen(2017)]{shen2017}
Z.~Shen.
\newblock Adaptive local parametric estimation of crop yields: implications for crop insurance ratemaking.
\newblock \emph{European Review of Agricultural Economics}, 45\penalty0 (2):\penalty0 173--203, 2017.

\bibitem[Sklar(1959)]{sklar1959}
A.~Sklar.
\newblock Fonctions de répartition à $n$ dimensions et leurs marges.
\newblock \emph{Publ. Inst. Statist. Univ. Paris 8}, pages 229--231, 1959.

\bibitem[{United Nations}(2015)]{UnitedNations2015}
{United Nations}.
\newblock Transforming our world: The 2030 agenda for sustainable development, 2015.
\newblock URL \url{https://sdgs.un.org/2030agenda}.

\bibitem[Ver{\'o}n et~al.(2015)]{veron2015}
S.~R. Ver{\'o}n et~al.
\newblock Climate impact on agricultural productivity in the pampas region of argentina.
\newblock \emph{Agricultural Systems}, 138:\penalty0 56--64, 2015.

\bibitem[Wang et~al.(2019)Wang, Yu, Yang, Yang, Gao, and Wang]{wang2019}
L.~Wang, H.~Yu, M.~Yang, R.~Yang, R.~Gao, and Y.~Wang.
\newblock A drought index: The standardized precipitation evapotranspiration runoff index.
\newblock \emph{Journal of Hydrology}, 571:\penalty0 651--668, 2019.

\bibitem[Zhu(2018)]{zhu2018}
X.~Zhu.
\newblock A dynamic factor approach for improved index insurance design and yield estimation.
\newblock \emph{SSRN Electronic Journal}, 2018.

\end{thebibliography}

\clearpage
\appendix

\section{ETCCDI Climate Extreme Indices}
\label{app:ETCCDI}
\begin{table}[h!]
\centering
\caption{ETCCDI Climate Extreme Indices}
\scalebox{0.8}{\begin{tabular}{llp{13cm}}
\toprule
\textbf{Category} & \textbf{Index Name} & \textbf{Description} \\
\midrule
\multirow{16}{*}{Temperature-Based} 
& \texttt{TXx} & Annual maximum of daily maximum temperatures. \\
& \texttt{TNx} & Annual maximum of daily minimum temperatures. \\
& \texttt{TXn} & Annual minimum of daily maximum temperatures. \\
& \texttt{TNn} & Annual minimum of daily minimum temperatures. \\
& \texttt{SU} & Number of summer days (days with maximum temperature $>$ 25°C). \\
& \texttt{TR} & Number of tropical nights (days with minimum temperature $>$ 20°C). \\
& \texttt{FD} & Number of frost days (days with minimum temperature $<$ 0°C). \\
& \texttt{ID} & Number of icing days (days with maximum temperature $<$ 0°C). \\
& \texttt{GSL} & Growing season length (annual count of days between first span of at least 6 days with daily mean temperature > 5°C and first span after July 1st of 6 days with daily mean temperature $<$ 5°C). \\
& \texttt{DTR} & Daily temperature range (monthly mean difference between daily maximum and minimum temperature). \\
& \texttt{TN10p} & Percentage of days when daily minimum temperature $<$ 10th percentile. \\
& \texttt{TX10p} & Percentage of days when daily maximum temperature $<$ 10th percentile. \\
& \texttt{TN90p} & Percentage of days when daily minimum temperature $>$ 90th percentile. \\
& \texttt{TX90p} & Percentage of days when daily maximum temperature $>$ 90th percentile. \\
& \texttt{WSDI} & Warm spell duration index (annual count of days with at least 6 consecutive days when daily maximum temperature $>$ 90th percentile). \\
& \texttt{CSDI} & Cold spell duration index (annual count of days with at least 6 consecutive days when daily minimum temperature $<$ 10th percentile). \\
\midrule
\multirow{11}{*}{Precipitation-Based} 
& \texttt{PRCPTOT} & Annual total precipitation in wet days (days with precipitation $\geq$ 1mm). \\
& \texttt{SDII} & Simple daily intensity index (ratio of annual total precipitation to the number of wet days). \\
& \texttt{R10mm} & Number of heavy precipitation days (days with precipitation $\geq$ 10mm). \\
& \texttt{R20mm} & Number of very heavy precipitation days (days with precipitation $\geq$ 20mm). \\
& \texttt{Rnnmm} & Number of days with precipitation $\geq$ nnmm (user-defined threshold). \\
& \texttt{RX1day} & Annual maximum 1-day precipitation. \\
& \texttt{RX5day} & Annual maximum consecutive 5-day precipitation. \\
& \texttt{CDD} & Consecutive dry days (maximum number of consecutive days with precipitation $<$ 1mm). \\
& \texttt{CWD} & Consecutive wet days (maximum number of consecutive days with precipitation $\geq$ 1mm). \\
& \texttt{R95pTOT} & Annual total precipitation from very wet days (days with precipitation $>$ 95th percentile). \\
& \texttt{R99pTOT} & Annual total precipitation from extremely wet days (days with precipitation $>$ 99th percentile). \\
\bottomrule
\end{tabular}}
\label{tab:etccdi_indices}
\end{table}

\section{Correlation and tail-dependence matrices for the Core Climate Extreme Indices}
\label{app:1}
\begin{figure}[h!]
\centering
\includegraphics[width=1\textwidth]{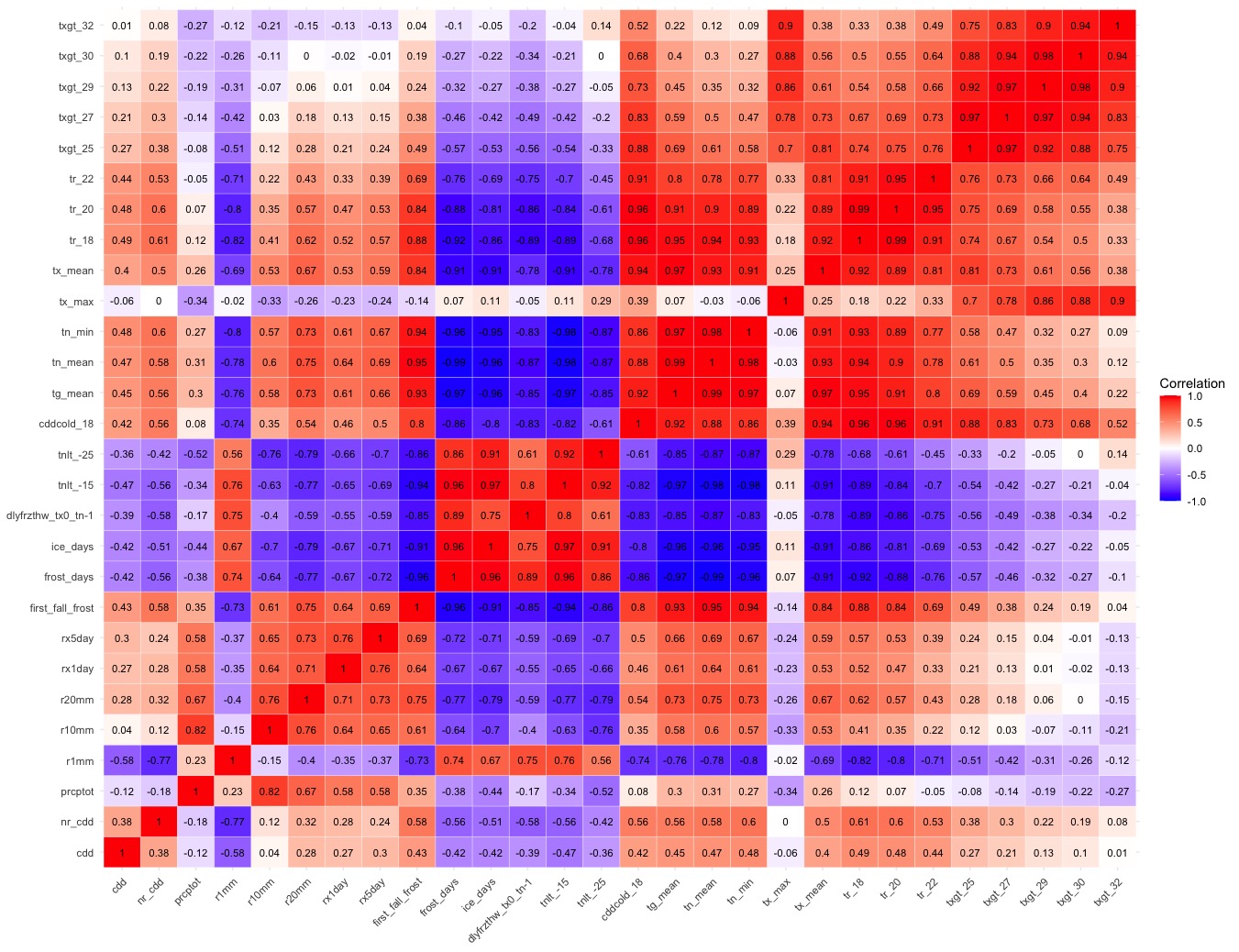}
\caption{Correlation heat map of indices.}
\label{fig:map3}
\end{figure}

\begin{figure}[h!]
\centering
\includegraphics[width=1\textwidth]{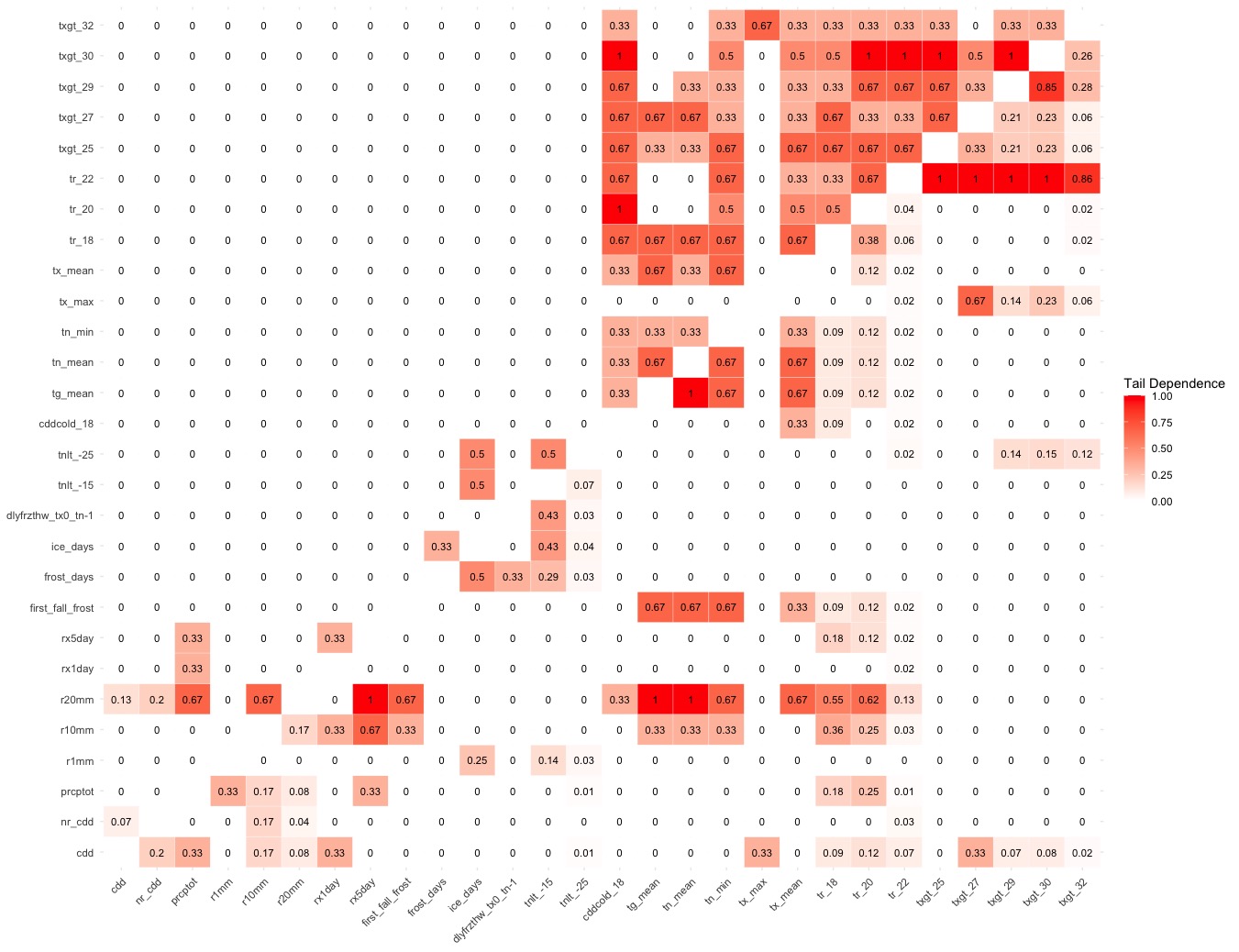}
\caption{1\% tail dependence of indices.}
\label{fig:map§}
\end{figure}

\begin{figure}[h!]
\centering
\includegraphics[width=1\textwidth]{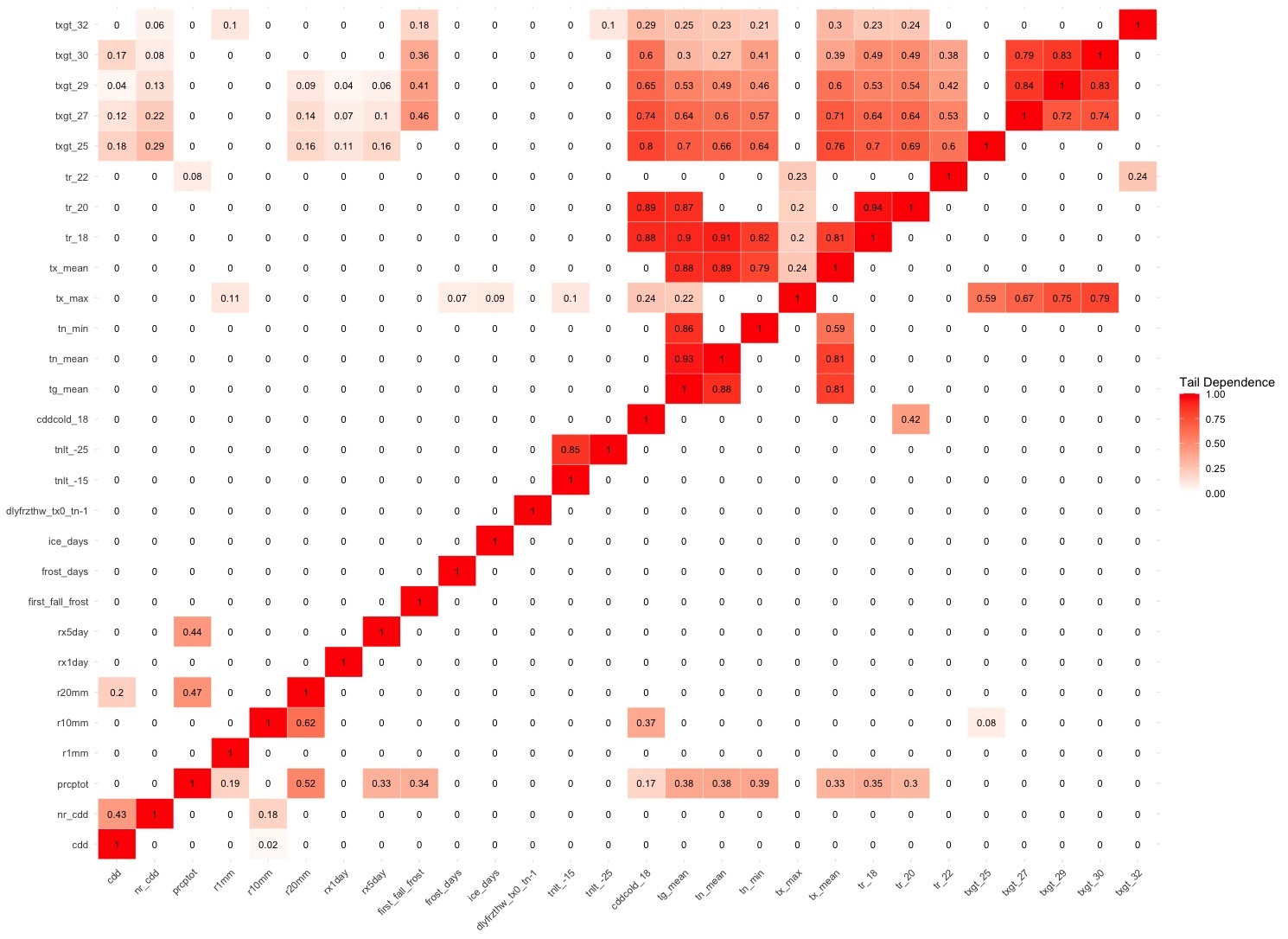}
\caption{Upper and lower tail dependence coefficients.}
\label{fig:map4}
\end{figure}

\clearpage
\section{Parameter estimates for the four-dimensional copula model}
\label{app:2}

\begin{table}[h!]
\centering
\caption{Estimated parameters for the conditional copula}
\begin{tabular}{lccccccc}
\toprule
Copula & $\omega$ & $\alpha$ & $\gamma_1$ & $\gamma_2$ & $\gamma_3$ & $\gamma_4$ & $\gamma_5$ \\
\midrule
Gaussian & 0.8191 & 0.4949 & 0.0191 & -0.0092 & -0.0034 & 0.0191 & 0.0122 \\
Clayton  & 0.8016 & 0.5029 & -0.0015 & 0.0018 & -0.0011 & 0.0014 & -0.0001 \\
Frank    & 0.8173 & 0.4834 & 0.0173 & 0.4079 & 0.0173 & 0.0173 & 0.0173 \\
Gumbel   & 0.8096 & 0.4905 & 0.0092 & -0.0791 & 0.0034 &  0.0164 & 0.0142\\
Joe      & 0.8190 & 0.4908 & 0.0190 & 0.0190 & 0.0190 & 0.0190 & 0.0190 \\
\bottomrule
\end{tabular}
\label{tab:copulaparam}
\end{table}

\begin{sidewaystable}
\begin{center}
\begin{minipage}{\textwidth}
\caption{Estimated parameters for the dynamic GEV}
\scalebox{0.7}{
\begin{tabular}{llcccccccccccccccccccc}
\toprule
\multirow{2}{*}{Region $(d)$} & \multirow{2}{*}{Copula}  & \multicolumn{4}{c}{$Z_{d,1,t}$} & \multicolumn{4}{c}{$Z_{d,2,t}$}& \multicolumn{4}{c}{$Z_{d,3,t}$} & \multicolumn{4}{c}{$Z_{d,4,t}$} & \multicolumn{4}{c}{$Z_{d,5,t}$} \\ \cline{3-22}
& & $\phi_{d,1}$ & $\sigma_{\mu_{d,1}}$ & $\sigma_{d,1}$ & $\xi_{d,1}$ &$\phi_{d,2}$ & $\sigma_{\mu_{d,2}}$ & $\sigma_{d,2}$ & $\xi_{d,2}$ & $\phi_{d,3}$ & $\sigma_{\mu_{d,3}}$ & $\sigma_{d,3}$ & $\xi_{d,3}$ & $\phi_{d,4}$ & $\sigma_{\mu_{d,4}}$ & $\sigma_{d,4}$ & $\xi_{d,4}$ & $\phi_{d,5}$ & $\sigma_{\mu_{d,5}}$ & $\sigma_{d,5}$ & $\xi_{d,5}$  \\
\midrule
\multirow{4}{*}{Dufferin} 
& Gaussian & -0.0016 & -0.0005 & 7.2421 & 7.2472 & 0.0058 & 1.0770 & 1.0791 & 0.0094 
& 0.0007 & 8.2467 & 7.8175 & 0.0095 & -0.0082 & 0.5551 & 0.5588 & 0.0095&0.0000 & 3.2420 & 3.2474 
& 0.0091 \\
& Clayton & 0.0000 & 0.0000 & 7.2414 & 7.2414 & 0.0000 & 1.0769 & 1.0769 & 0.0100 
& 0.0000 & 7.8117 & 7.8117 & 0.0100 & 0.0000 & 0.5582 & 0.5582 &0.0100 &0.0000 & 3.2478 & 3.2478 
& 0.0100 \\
& Frank & 0.4007 & 0.0114 & 7.1496 & 7.2503 & -0.0025 & 1.0797 & 1.0880 & 0.0094 
& 0.0114 & 7.8231 & 7.8137 & 0.0153 & 0.0114 & 0.5581 & 0.5696 & 0.0166&0.0166 & 3.2592 & 3.2496 
& 0.0171 \\
& Gumbel & 0.0197 & 0.0197 & 7.2611 & 7.2611 & 0.0197 & 1.0756 & 1.0916 & 0.0143 
& -0.0152 & 7.8177 & 7.8313 & -0.0002 & 0.0197 & 0.5578 & 0.5778 &0.0005 &0.0005 & 3.2675 & 3.2675 
& 0.0200 \\
& Joe & 0.0190 & 0.0163 & 7.2521 & 7.2678 & 0.0264 & 1.0609 & 1.1033 & 0.0088 
& -0.0023 & 7.8381 & 7.8317 & -0.0138 & 0.0264 & 0.5424 & 0.5846 &0.0254 &0.0254 & 3.2742 & 3.2742 
& -0.0001 \\
\midrule 
\multirow{4}{*}{Wellington} 
& Gaussian & 0.0083 & -0.0013 & 4.8799 & 4.8799 & 0.0258 & 1.0240 & 1.0310 & 0.0074 
& -0.0067 & 6.7443 & 6.7406 & 0.0062 & 0.0019 & 0.3801 & 0.3829 & 0.0118 & 0.0058 & 3.2943 
& 3.2984 & 0.0110\\
& Clayton & 0.0100 & 0.0000 & 4.8741 & 4.8741 & 0.0000 & 1.0252 & 1.4158 & 0.0100 
& 0.0000 & 6.7429 & 6.7429 & 0.0100 & 0.0000 & 0.3804 & 0.3804 & 0.0100 & 0.0000 & 3.2996 
& 3.2996 & 00100\\
& Frank & 0.0106 & 0.0114 & 4.8689 & 4.8785 & 0.0114 & 1.0239 & 1.0366 & 0.0002 
& -0.0045 & 6.7543 & 6.7533 & 0.0162 & 0.0066 & 0.3918 & 0.5871 & 0.0103 & 0.0067 & 3.3110 
& 3.3110& 0.0163 \\
& Gumbel & 0.0130 & 0.0197 & 4.8809 & 4.8937 & 0.0197 & 1.0300 & 1.0448 & 0.0015 
& 0.0073 & 7.1531 & 6.7591 & 0.0204 & 0.0111 & 0.3322 & 0.4000 & 0.0197 & 0.0114 & 3.3192 
& 3.3192& 0.0114 \\
& Joe & 0.0093 & 0.0069 & 4.9249 & 4.8836 & 0.0170 & 1.0109 & 1.0449 & 0.0013 
& 0.0202 & 6.7660 & 6.7692 & 0.0196 & -0.0860 & 0.3407 & 0.3304 & 0.0264 & 0.0213 & 3.3260 
& 3.3260& 0.0213\\
\bottomrule
\end{tabular}}
\label{tab:gevparam}
\end{minipage}
\end{center}
\end{sidewaystable}

\end{document}


\maketitle

\subsubsection*{Example: Clayton copula, $D=2$, $M=1$} 
Suppose that we have only $D=2$ regions and only consider the maximum level of precipitation observed during the year as covariate. We have the following random variables:
\begin{itemize}
    \item $Y_{d,t}$ for $d=1,2$ and $t=1,\ldots,T$: crop yield in region $d$ at time $t$
    \item $Z_{d,t}$: maximum amount of precipitation reported in region $d$ at time $t$
    \item $X_{t} = \max(Z_{1,t},Z_{2,t})$: maximum amount of precipitation observed across both regions
\end{itemize}
\paragraph{Marginal distributions} 
We fit the marginal distributions for the crop yields using Bayesian Structural Time Series with a local linear trend, one covariate that we include dynamically in the observation equation and one auto-regressive component: 
\begin{align*}
    y_{1,t} &= l_{1,t} + \psi_{1} e_{1,t-1} + \beta_{1,t} Z_{1,t} + \epsilon_{1,t} \\
    y_{2,t} &= l_{2,t} + \psi_{2} e_{2,t-1} + \beta_{2,t} Z_{2,t} + \epsilon_{2,t}.
\end{align*}
For $d=1,2$, we then have
\begin{align*}
    l_{d,t} &= l_{d,t-1} + \tau_{d,t-1} + \nu_{d,t}, \hspace{0.3cm}\nu_{d,t}\sim\mathcal{N}(0,\sigma^2_{\nu_d}) \\
    \tau_{d,t} &= \tau_{d,t-1} + \zeta_{d,t}, \hspace{0.3cm}\zeta_{d,t}\sim\mathcal{N}(0,\sigma^2_{\zeta_d}) \\
    e_{d,t} &= \psi_{d} e_{d,t-1} + \eta_{d,t}, \hspace{0.3cm}\eta_{d,t}\sim\mathcal{N}(0,\sigma^2_{\eta_d}) \\
    \beta_{d,t} &= \beta_{d,t-1} + \lambda_{d,t}, \hspace{0.3cm}\lambda_{d,t}\sim\mathcal{N}(0,\sigma^2_{\lambda_d}) \\
    \epsilon_{d,t} &\sim \mathcal{N}(0,\sigma^2_{\epsilon_d}).
\end{align*}
We specify prior distributions for the 10 variance parameters, as well as for the 2 auto-regressive coefficients. We choose the commonly used Inverse-Gamma priors for the variance parameters. The priors set on $\psi_1$ and $\psi_2$ must ensure stationarity of the AR(1) process. A typical choice is then the uniform distribution. The full list of prior distributions is given hereafter for $d=1,2$:
\begin{itemize}
    \item $\psi_d \sim \mathcal{U}(-1, 1)$
    \item $\sigma^2_{\epsilon_d}\sim \text{Inverse-Gamma}(\alpha_{\epsilon_d}, \beta_{\epsilon_d})$
     \item $\sigma^2_{\nu_d}\sim \text{Inverse-Gamma}(\alpha_{\nu_d}, \beta_{\nu_d})$
     \item $\sigma^2_{\zeta_d}\sim \text{Inverse-Gamma}(\alpha_{\zeta_d}, \beta_{\zeta_d})$
     \item $\sigma^2_{\eta_d}\sim \text{Inverse-Gamma}(\alpha_{\eta_d}, \beta_{\eta_d})$
     \item $\sigma^2_{\lambda_d}\sim \text{Inverse-Gamma}(\alpha_{\lambda_d}, \beta_{\lambda_d})$
\end{itemize}
The posterior distribution in each region is thus given by
\begin{align*}
    f(\psi_d, \sigma_{\epsilon_d}, \sigma_{\nu_d},\sigma_{\zeta_d},\sigma_{\eta_d},\sigma_{\lambda_d}\mid \vect{y}_d) \propto &\Bigg[\prod_{t=1}^{T} \frac{1}{\sqrt{2\pi\sigma^2_{\epsilon_d}}}\exp \Bigg(-\frac{\epsilon_{d,t}^2}{2\sigma^2_{\epsilon_d}}\Bigg) \Bigg] \\
    & \times \frac{1}{2}\\
    & \times \frac{(\beta_{\epsilon_d})^{\alpha_{\epsilon_d}}}{\Gamma(\alpha_{\epsilon_d})} (1/\sigma_{\epsilon_d})^{(\alpha_{\epsilon_d} + 1)} \exp(-\beta_{\epsilon_d}/\sigma_{\epsilon_d}) \\
    & \times \frac{(\beta_{\nu_d})^{\alpha_{\nu_d}}}{\Gamma(\alpha_{\nu_d})} (1/\sigma_{\nu_d})^{(\alpha_{\nu_d} + 1)} \exp(-\beta_{\nu_d}/\sigma_{\nu_d}) \\
    & \times \frac{(\beta_{\zeta_d})^{\alpha_{\zeta_d}}}{\Gamma(\alpha_{\zeta_d})} (1/\sigma_{\zeta_d})^{(\alpha_{\zeta_d} + 1)} \exp(-\beta_{\zeta_d}/\sigma_{\zeta_d}) \\
    & \times \frac{(\beta_{\eta_d})^{\alpha_{\eta_d}}}{\Gamma(\alpha_{\eta_d})} (1/\sigma_{\eta_d})^{(\alpha_{\eta_d} + 1)} \exp(-\beta_{\eta_d}/\sigma_{\eta_d}) \\
    & \times \frac{(\beta_{\lambda_d})^{\alpha_{\lambda_d}}}{\Gamma(\alpha_{\lambda_d})} (1/\sigma_{\lambda_d})^{(\alpha_{\lambda_d} + 1)} \exp(-\beta_{\lambda_d}/\sigma_{\lambda_d}).
\end{align*}
\paragraph{Clayton copula} The Clayton copula is written as
\begin{align*}
    C(u_{1,t}, u_{2,t}\mid \theta_t(\vect{X}_t)) = \Big[ u_{1,t}^{-\theta_t(\vect{X}_t)} + u_{2,t}^{-\theta_t(\vect{X}_t)} - 1 \Big]^{-1/\theta_t(\vect{X}_t)},
\end{align*}
with
\begin{align*}
    \theta_t(\vect{X}_t) = \exp(\omega + \alpha_t \theta_{t-1} + \gamma_t X_{t} + \epsilon_t).
\end{align*}
The choice of the exponential link function allows to maintain $\theta_t(\vect{X}_t)$ in $\mathbb{R}^+$. The copula density is then
\begin{align*}
    c(u_{1,t}, u_{2,t} \mid \theta_t(\vect{X}_t)) = (\theta_t(\vect{X}_t) + 1)u_{1,t}^{-\theta_t(\vect{X}_t) - 1} u_{2,t}^{-\theta_t(\vect{X}_t) - 1}(u_{1,t}^{-\theta_t(\vect{X}_t)} + u_{2,t}^{-\theta_t(\vect{X}_t)} - 1)^{-\frac{2\theta_t(\vect{X}_t) + 1}{\theta_t(\vect{X}_t)}}.
\end{align*}
\paragraph{Dynamic GEV distributions} We use again dynamic GEV distributions to model $Z_{1}$ and $Z_2$ (and, as such, $X$). The density of $Z_{d}$ for $d=1,2$, conditionnally on the time-varying parameters $\mu_{d,t}$ for $d=1,2$, is given by
\begin{align*}
    f_{Z_{d}}(z_{d,t} \mid \mu_{d,t}, \sigma_d,\xi_d) = \frac{1}{\sigma_{d}}\Bigg[1 + \xi_{d} \frac{z_{d,t}-\mu_{d,t}}{\sigma_{d}}\Bigg]^{-1-1/\xi_{d}} \exp\Bigg\{-\Big[1 + \xi_{d} \frac{z_{d,t}-\mu_{d,t}}{\sigma_{d}}\Big]^{-1/\xi_{d}}\Bigg\},
\end{align*}
with 
\begin{align*}
    \mu_{d,t} = \phi_d \mu_{d,t-1} + \epsilon_{d,t}, \hspace{0.3cm} \epsilon_{d,t}\sim\mathcal{N}(0,\sigma^2_{\mu_d}).
\end{align*}
\paragraph{Full model} Combining the expressions derived above, the full model pseudo-likelihood in this example can be written as 
\begin{align*}
    \mathcal{L}_{\text{full}}(\vect{\Theta}) &= \mathcal{L}_{\text{copula}}(\vect{\alpha}, \vect{\gamma}, \vect{\sigma}_{\theta}) \times \mathcal{L}_{\text{GEV}}(\vect{\phi}, \vect{\sigma_{\mu}}, \vect{\sigma}, \vect{\xi}) \\
    & = \prod_{t=1}^{T} \Bigg\{ (\theta_t(\vect{X}_t) + 1)u_{1,t}^{-\theta_t(\vect{X}_t) - 1} u_{2,t}^{-\theta_t(\vect{X}_t) - 1}(u_{1,t}^{-\theta_t(\vect{X}_t)} + u_{2,t}^{-\theta_t(\vect{X}_t)} - 1)^{-\frac{2\theta_t(\vect{X}_t) + 1}{\theta_t(\vect{X}_t)}} \\
    & \hspace{1.5cm} \times \prod_{d=1}^{2} \Big[\frac{1}{\sigma_{d}}\Big[1 + \xi_{d} \frac{z_{d,t}-\mu_{d,t}}{\sigma_{d}}\Big]^{-1-1/\xi_{d}} \exp\Big\{-\Big[1 + \xi_{d} \frac{z_{d,t}-\mu_{d,t}}{\sigma_{d}}\Big]^{-1/\xi_{d}}\Big\} \\
    & \hspace{2.5cm}\times \frac{1}{\sqrt{2\pi}\sigma_{\mu_{d}}}\exp\Big(-\frac{(\mu_{d,t} - \phi_{d}\mu_{d,t-1})^2}{2\sigma^2_{\mu_{d}}}\Big) \Big]\Bigg\}.
\end{align*}